\documentclass[11pt,a4paper]{article}
\usepackage{jheppub}
\allowdisplaybreaks

\def\be{\begin{equation}}
\def\ee{\end{equation}}
\def\bea{\begin{eqnarray}}
\def\eea{\end{eqnarray}}
\def\beal{\begin{equation}\begin{aligned}}
\def\eeal{\end{aligned}\end{equation}}
\def\nn{\nonumber}

\def\bra#1{\langle #1|}
\def\ket#1{|#1 \rangle}

\def\braket#1{\langle #1 \rangle}

\def\u#1{\underline{#1}}
\def\o#1{\overline{#1}}

\def\la{\lambda}
\def\lb{\tilde{\lambda}}

\def\Res_#1{\operatorname*{Res}_{#1}}
\def\sgn{\operatorname*{sgn}}
\def\Tr{\operatorname*{Tr}}
\def\tr{\operatorname*{tr}}

\def\tf{\tilde{f}}

\def\ie{i.e. }
\def\eg{e.g. }
\def\etc{etc. }

\def\eqn#1{eq.~\eqref{#1}}
\def\eqns#1#2{eqs.~\eqref{#1} and~\eqref{#2}}
\def\eqnss#1#2#3{eqs.~\eqref{#1}, \eqref{#2} and~\eqref{#3}}

\def\fig#1{figure~{\ref{#1}}}

\def\sec#1{section~{\ref{#1}}}

\def\app#1{appendix~{\ref{#1}}}

\def\rcite#1{ref.~\cite{#1}}
\def\rcites#1{refs.~\cite{#1}}
\def\Rcite#1{Ref.~\cite{#1}}

\newcommand{\scalegraph}[2]{\vcenter{\hbox{\!\;\includegraphics[scale=#1]{graphs/#2.pdf}\!\;}}}

\title{Helicity amplitudes for QCD with massive quarks}

\author{Alexander Ochirov}

\affiliation{ETH Z\"urich, Institut f\"ur Theoretische Physik,
Wolfgang-Pauli-Str. 27, 8093 Z\"urich, Switzerland}

\emailAdd{aochirov@phys.ethz.ch}

\abstract{
The novel massive spinor-helicity formalism of Arkani-Hamed, Huang and Huang
provides an elegant way to calculate scattering amplitudes in
quantum chromodynamics for arbitrary quark spin projections.
In this note we compute two families
of tree-level QCD amplitudes with one massive quark pair and $n-2$ gluons.
The two cases include all gluons with identical helicity
and one opposite-helicity gluon being color-adjacent to one of the quarks.
Our results naturally incorporate the previously known amplitudes
for both quark spins quantized along one of the gluonic momenta.
In the all-multiplicity formulae presented here the spin quantization axes
can be tuned at will, which includes the case of the definite-helicity quark states.
}

\begin{document}
\maketitle

\section{Introduction}
\label{sec:intro}

The recent advances in the analytic understanding of the scattering amplitudes
are often believed to be specific to massless theories,
preferably with supersymmetry.
It is arguably due to the absence, until recently,
of a fully satisfactory spinor-helicity formalism for massive particles.
Of course, the massless spinor-helicity formalism~\cite{Berends:1981rb,DeCausmaecker:1981bg,Gunion:1985vca,Kleiss:1985yh,Xu:1986xb,Gastmans:1990xh}
(popularized \eg by \rcite{Dixon:1996wi})
has been applied~\cite{Kleiss:1986qc,Dittmaier:1998nn,Schwinn:2005pi}
to define massive Dirac spinors.
However,
that construction did not manage to dispel the notion of the on-shell amplitude methods
being restricted to the massless case.
Recently, however, Arkani-Hamed, Huang and Huang~\cite{Arkani-Hamed:2017jhn}
have introduced a complete version of a massive spinor-helicity formalism
and used it to reconsider an array of quantum field-theoretic results
from the fully on-shell perspective.

This note is about how this massive formalism can be used in one field theory of interest~---
quantum chromodynamics with heavy quarks.
For simplicity, here we only consider the amplitudes with one massive quark-antiquark pair,
with the other particles being gluons of definite helicity.
The main goals of this note are two-fold:
\begin{itemize}
\item We provide new all-multiplicity expressions, \eqns{QQggnAP}{QQggnOM},
for the $n$-point color-ordered amplitudes with two quarks
in case of all gluons of identical helicity
and the case of one gluon of opposite helicity color-adjacent to one of the quarks.
\item We pay special attention to our conventions
so that our results be consistent with the vast QCD literature.
That involves flexible transitions between the presented massive formalism,
its massless analogue recovered in the high-energy limit,
the general Dirac spinors and their realization using the massless Weyl spinors.
\end{itemize}

In view of the second goal, in \sec{sec:spinors} we review the spinor-helicity formalism 
in an effort to combine brevity with comprehensiveness.
We illustrate the introduced methods in \sec{sec:4pt},
where we show two ways to derive a full color-dressed amplitude
for four-particle scattering (corresponding \eg to non-abelian Compton scattering).
We highlight the difference between the Feynman-diagrammatic approach
and the on-shell construction, which deals solely with gauge-invariant quantities.

In \sec{sec:npt} we present and prove the aforementioned all-multiplicity amplitudes
with two specific gluon-helicity configurations. For that we employ
the Britto-Cachazo-Feng-Witten (BCFW) on-shell recursion~\cite{Britto:2004ap,Britto:2005fq}.
The spins of the quark and the antiquark remain unfixed throughout the calculations,
which lets us specialize to the specific quark-spin projections
considered previously~\cite{Schwinn:2007ee}
in the massless-spinor-based formalism~\cite{Kleiss:1986qc,Dittmaier:1998nn,Schwinn:2005pi}.
Hence, in \sec{sec:checks}, we give a simple dictionary~\eqref{massivespinors2old}
between the two descriptions and thus compare our results with the literature.
It also shows that the new formalism easily incorporates the old one,
the elegance of which suffered from
the loss of the explicit little-group ${\rm SU}(2)$ symmetry.

We hope that this note will pave the way to more tree- and loop-level
calculations in the newly complete spinor-helicity formalism~\cite{Arkani-Hamed:2017jhn},
as outlined in \sec{sec:outro}.

\section{Spinor-helicity review}
\label{sec:spinors}

It is well-known that
particles are defined as irreducible unitary representations
of the Poincare group~\cite{Wigner:1939cj,Bargmann:1948ck}.
Once the translation operator is diagonalized
and the particles are labeled by their momentum~$p^\mu$,
one is left with the Lorentz ${\rm SO}(1,3)$ subgroup of the Poincare group.
The remaining labels of a one-particle state turn out to
belong to a representation of its little group.
This subgroup of ${\rm SO}(1,3)$ is crucial for understanding spin.
It is defined through the Lorentz transformations
that preserve the momentum~$p^\mu$ of the particle.
It corresponds to ${\rm SO}(2)$ for massless states
or to ${\rm SO}(3)$ for massive ones.

To include fermions into consideration, one must generalize to
the universal covering group ${\rm SL}(2,\mathbb{C})$ of ${\rm SO}(1,3)$.
The homomorphism between these two groups is implemented by the spinor maps
\be
   p_{\alpha \dot{\beta}} = p_{\mu} \sigma_{\alpha \dot{\beta}}^{\mu} , \qquad \quad
   p^{\dot{\alpha} \beta} = p^{\mu} \bar{\sigma}_{\mu}^{\dot{\alpha} \beta} .
\ee
The Pauli matrices\footnote{We use
$\sigma^0 = \big(\begin{smallmatrix} 1 & 0 \\ 0 & 1 \end{smallmatrix}\big)$,
$\sigma^1 = \big(\begin{smallmatrix} 0 & 1 \\ 1 & 0 \end{smallmatrix}\big)$,
$\sigma^2 = \big(\begin{smallmatrix} 0 & -i \\ i & 0 \end{smallmatrix}\big)$,
$\sigma^3 = \big(\begin{smallmatrix} 1 & 0 \\ 0 & -1 \end{smallmatrix}\big)$,
as well as
 $ \epsilon^{\alpha \beta}
    = \big(\begin{smallmatrix} 0 & 1 \\ -1 & 0 \end{smallmatrix}\big) $ and
 $ \epsilon_{\alpha \beta}
    = \big(\begin{smallmatrix} 0 & -1 \\ 1 & 0 \end{smallmatrix}\big) $.}
$\sigma^{\mu} = (1,\sigma^1,\sigma^2,\sigma^3)$ and
$\bar{\sigma}^{\mu} = (1,-\sigma^1,-\sigma^2,-\sigma^3)$
here translate Lorentz transformations between the spinorial and vectorial languages:
\be
   L^{\mu}_{~\nu} = \frac{1}{2} \tr\!\big( \bar{\sigma}^{\mu}S\sigma_{\nu}S^{\dagger}
                                     \big) : \qquad \quad
   p_{\alpha \dot{\delta}} \to S_{\alpha}^{~\beta} p_{\beta \dot{\gamma}}
                             \left(S_{\delta}^{~\gamma}\right)^* \qquad \Rightarrow \qquad
   p^\mu \to L^{\mu}_{~\nu} p^\nu ,
\label{sl2lorentz}
\ee
for $L\in{\rm SO}(1,3)$ and $S\in{\rm SL}(2,\mathbb{C})$.
At the same time, the little groups for massless and massive particles
are accordingly promoted to ${\rm U}(1)$ and ${\rm SU}(2)$.

An important property of the ${\rm SL}(2,\mathbb{C})$ transformations
(and hence the ${\rm SU}(2)$ ones) is that they preserve the antisymmetric form $\epsilon_{\alpha\beta}=-\epsilon^{\alpha\beta}$,
\ie the spinor product:
\be
   S_{\alpha}^{~\gamma} S_{\beta}^{~\delta} \epsilon_{\gamma \delta} =
      \frac{1}{2} \epsilon_{\gamma \delta}
      ( \delta_{\alpha}^{\phi} \delta_{\beta}^{\psi}
      - \delta_{\beta}^{\phi} \delta_{\alpha}^{\psi} )
      S_{\phi}^{~\gamma} S_{\psi}^{~\delta} =
      \frac{1}{2} \epsilon_{\gamma \delta}
      \epsilon_{\alpha \beta} \epsilon^{\phi \psi}
      S_{\phi}^{~\gamma} S_{\psi}^{~\delta} =
      \epsilon_{\alpha \beta} \det S = \epsilon_{\alpha \beta} .
\label{sl2symplectic}
\ee
This form allows to raise and lower both the spinor and massive-little-group indices at will.

Now let us explore different spinor types one by one.
The massless and massive Weyl spinors comprise the spinor-helicity formalism~\cite{Berends:1981rb,DeCausmaecker:1981bg,Gunion:1985vca,Kleiss:1985yh,Xu:1986xb,Gastmans:1990xh,Arkani-Hamed:2017jhn},
while the Dirac spinors are helpful to connect it to the more traditional approaches.

\subsection{Massless Weyl spinors}

In the massless case, the on-shell condition $p^2=\det\{p_{\alpha\dot{\beta}}\}=0$ means
that the degenerate matrix $p_{\alpha \dot{\beta}}$ can be decomposed as a tensor product
of two Weyl spinors.
That decomposition can be written in various interchangeable ways
using the spinor bra-ket notation:
\be
   \begin{aligned}
   p_{\alpha \dot{\beta}} & = \la_{p\:\!\alpha} \lb_{p\:\!\dot{\beta}}
      \equiv \ket{p}_{\alpha} [p|_{\dot{\beta}} \\
   p^{\dot{\alpha} \beta} & \overset{\substack{\Updownarrow \\ ~}}{=}
                            \lb_p^{\dot{\alpha}} \la_p^{\beta}
      \equiv |p]^{\dot{\alpha}} \bra{p}^{\beta}
   \end{aligned}
   \qquad \quad \Leftrightarrow \qquad \quad
   \begin{aligned}
   \not{\!p}\,&= \ket{p} [p| + |p] \bra{p} \\
   p^{\mu} & \overset{\substack{\Updownarrow \\ ~}}{=}
      \frac{1}{2} \la_p^{\alpha} \sigma^{\mu}_{\alpha\dot{\beta}} \lb_p^{\dot{\beta}}
      \equiv \frac{1}{2} \bra{p}\sigma^{\mu}|p] \\
   p_{\mu} & \overset{\substack{\Updownarrow \\ ~}}{=}
      \frac{1}{2} \lb_{p\:\!\dot{\alpha}}
      \bar{\sigma}_{\mu}^{\dot{\alpha}\beta} \la_{p\:\!\beta}
      \equiv \frac{1}{2} [p|\bar{\sigma}_{\mu}\ket{p} .
   \end{aligned}
\label{momentumbispinor}
\ee
This notation fits
the spinor products~\cite{Berends:1981rb,DeCausmaecker:1981bg,
Gunion:1985vca,Kleiss:1985yh,Xu:1986xb,Gastmans:1990xh}
particularly well:
\be\!
   \braket{p\;\!q} \equiv \la_p^{\alpha} \la_{q\:\!\alpha}
    = \la_p^{\alpha} \epsilon_{\alpha\beta} \la_q^{\beta} , \qquad \quad
   [p\;\!q] \equiv \lb_{p\:\!\dot{\alpha}} \lb_q^{\dot{\alpha}}
    = \lb_{p\dot{\alpha}} \epsilon^{\dot{\alpha}\dot{\beta}}
      \lb_{q\:\!\dot{\beta}} , \qquad \quad
   \braket{p\;\!q} [q\;\!p] = 2\;\!p\:\!\!\cdot\:\!\!q .\!
\label{spinorproducts}
\ee
The Lorentz transformations~\eqref{sl2lorentz} act on the Weyl spinors
$\la_{p\:\!\alpha} \equiv \ket{p}_\alpha$ and $\lb_p^{\dot{\alpha}}\equiv|p]^{\dot{\alpha}}$
via $S \in {\rm SL}(2,\mathbb{C})$,
but only up to the little-group ${\rm U}(1)$ rotations:\footnote{In the case that
the Lorentz transformation $L$ is a pure ${\rm SO}(2)$ rotation
around the momentum axis $\hat{p}$ by the angle $\phi$,
the little-group phases in \eqn{lorentz2littlegroupmassless} are unambiguous
and precisely equal to $\pm \phi/2$.}
\beal
   \la_{p\:\!\alpha} & ~\to~ S_{\alpha}^{~\beta} \la_{p\:\!\beta}
    ~=~ e^{i\phi/2} \la_{Lp\:\!\alpha} , \qquad \qquad~~\:\!
   \la_{p}^{\alpha} ~\to~ \la_{p}^{\beta} (S^{-1})_{\beta}^{~\alpha}
    ~=~ e^{i\phi/2} \la_{Lp}^{\alpha} , \\
   \lb_{p\:\!\dot{\alpha}} & ~\to~
      \lb_{p\:\!\dot{\beta}} (S^{\dagger})^{\dot{\beta}}_{~\dot{\alpha}}
    ~=~ e^{-i\phi/2} \lb_{Lp\:\!\dot{\alpha}} , \qquad \quad
   \lb_{p}^{\dot{\alpha}} ~\to~
      (S^{\dagger\;\!-1})^{\dot{\alpha}}_{~\dot{\beta}} \lb_{p}^{\dot{\beta}}
    ~=~ e^{-i\phi/2} \lb_{Lp}^{\dot{\alpha}} .
\label{lorentz2littlegroupmassless}
\eeal
These spinors also give us the building blocks
for the polarization vectors of gauge bosons:\!\!
\begin{subequations} \begin{align}
   \varepsilon_{p+}^{\mu} & =  \frac{1}{\sqrt{2}}
      \frac{\bra{q}\sigma^{\mu}|p]}{\braket{q\;\!p}}
   \qquad \quad~\:\: \Leftrightarrow \qquad \quad
   \not{\;\!\!\varepsilon}_{p}^{+} = \sqrt{2}\;\!
      \frac{\ket{q}[p|+|p]\bra{q}}{\braket{q\;\!p}} , \\
   \varepsilon_{p-}^{\mu} & = -\frac{1}{\sqrt{2}}
      \frac{[q|\bar{\sigma}^{\mu}\ket{p}}{[q\;\!p]}
   \qquad \quad \Leftrightarrow \qquad \quad
   \not{\;\!\!\varepsilon}_{p}^{-} = -\sqrt{2}\;\!
      \frac{\ket{p}[q|+|q]\bra{p}}
           {[q\;\!p]} ,
\end{align} \label{polvectors}%
\end{subequations}
where $q$ can be any null vector such that
$\ket{q}\,{\not \sim}\,\ket{p}$ and $|q]\,{\not \sim}\,|p]$.
Indeed, different reference vectors are equivalent up to a pure gauge, \eg
\be
   \varepsilon_{p+}^{\mu}(q') = \varepsilon_{p+}^{\mu}(q)
    + \frac{\sqrt{2} \braket{q' q} p^\mu}
           {\braket{q' p} \braket{p\;\!q}} .
\ee
Now it is important to note that
under a Lorentz transformation~\eqref{lorentz2littlegroupmassless}
the polarization vectors do not actually transform as proper vectors.
For instance, comparing
\be\!\!\!
   L^{\mu}_{~\nu} \varepsilon_{p+}^{\nu}
    = \frac{\bra{\la_q S^{-1}}\sigma^{\mu}|S^{\dagger\;\!-1}\lb_p]}
           {\sqrt{2}\braket{\la_q S^{-1}|S\la_p}}
    = e^{-i\phi} \frac{\bra{\la_q S^{-1}}\sigma^{\mu}|\lb_{Lp}]}
                {\sqrt{2}\braket{\la_q S^{-1}|\la_{Lp}}} \qquad \text{vs.} \qquad
   \varepsilon_{Lp+}^{\mu}
    \equiv \frac{\bra{\la_q}\sigma^{\mu}|\lb_{Lp}]}
                {\sqrt{2}\braket{\la_q\;\!\la_{Lp}}} ,\!
\ee
we conclude that Lorentz transformations act as
\be
   \varepsilon_{p\pm}^{\mu} ~\to~
   L^{\mu}_{~\nu} \varepsilon_{p\pm}^{\nu} ~\sim~ e^{\mp i\phi} \varepsilon_{Lp\pm}^{\mu}
\label{lorentz2littlegroupvector}
\ee
only up to an additional term
proportional to the new momentum $L^{\mu}_{~\nu} p^\nu$.
However, up to this caveat, this shows that these polarization vectors can be thought of as
conversion coefficients between the off-shell Lorentz transformations
and the corresponding on-shell little-group rotations~\cite{Arkani-Hamed:2017jhn}.
A similar statement for the Weyl spinors is demonstrated by \eqn{lorentz2littlegroupmassless}
and is also true for the massive case, see \eqn{lorentz2littlegroupmassive} below.

As a concrete realization of the Weyl spinors, one could use, for instance,
\be
   \la_{p\:\!\alpha} = \sqrt{2E}
      \begin{pmatrix}\!-e^{-i\varphi}\sin\frac{\theta}{2} \\
      \cos\frac{\theta}{2} \end{pmatrix}\!, \qquad \quad
   \lb_{p}^{\dot{\alpha}} = \sqrt{2E}
      \begin{pmatrix} \cos\frac{\theta}{2} \\
      e^{i\varphi}\sin\frac{\theta}{2} \end{pmatrix}\!,
\label{spinorsolutionangles}
\ee
for a null momentum expressible as
$p^\mu=E(1,\cos\varphi \sin\theta,\sin\varphi \sin\theta,\cos\theta)$.
A more practical implementation is given in \app{sec:masslessspinors}.

\subsection{Massive Weyl spinors}

For a nonzero mass $m$, we have a non-degenerate matrix~$p_{\alpha\dot{\beta}}$
that satisfies $\det\{p_{\alpha\dot{\beta}}\}=m^2$.
The Weyl spinors are then introduced~\cite{Arkani-Hamed:2017jhn}
by expanding $p_{\alpha\dot{\beta}}$ in terms of two explicitly degenerate matrices
$\la_{p\:\!\alpha}^{~\;1} \lb_{p\:\!\dot{\beta}1}$ and
$\la_{p\:\!\alpha}^{~\;2} \lb_{p\:\!\dot{\beta}2}$:
\beal
   p_{\alpha\dot{\beta}} & = \la_{p\:\!\alpha}^{~\;a} \lb_{p\:\!\dot{\beta}a}\!
    = \la_{p\:\!\alpha}^{~\;a} \epsilon_{ab} \lb_{p\:\!\dot{\beta}}^{~b} , & \qquad \quad
   \det\{\la_{p\:\!\alpha}^{~\;a}\} & = m , & \qquad \quad
   \det\{\lb_{p\:\!\dot{\alpha}}^{\;~a}\} & = m , \\ \!\!
   p^{\dot{\alpha}\beta} & \overset{\substack{\Updownarrow \\ ~}}{=}
                             \lb_{p\:\!a}^{\dot{\alpha}} \la_p^{\beta a}\!
    =-\lb_{p}^{\dot{\alpha} a} \epsilon_{ab} \la_p^{\beta b} , & \qquad \quad
   \la_{\alpha a} \la^{\beta a} & \overset{\substack{\Downarrow \\ ~}}{=}
      m \delta_\alpha^\beta , & \qquad \quad
   \lb^{\dot{\alpha} a} \lb_{\dot{\beta} a} & \overset{\substack{\Downarrow \\ ~}}{=}
      m \delta^{\dot{\alpha}}_{\dot{\beta}} .\!
\label{momentumbispinormassive}
\eeal
Here we have already indicated that the little-group indices $a,b=1,2$
are lowered and raised by the antisymmetric form $\epsilon_{ab}$,
preserved by ${\rm SU}(2)$ rotations.
Such little-group transformations
follow from the action of the Lorentz group on these spinors:
\beal
   \la_{p\:\!\alpha}^{~\;a} & ~\to~ S_{\alpha}^{~\beta} \la_{p\:\!\beta}^{~a}
    ~=~ \omega^a_{~b} \la_{Lp\:\!\alpha}^{~~\;b} , \qquad \quad~~~\;
   \la_{p}^{\alpha a} ~\to~ \la_{p}^{\beta a} (S^{-1})_{\beta}^{~\alpha}
    ~=~ \omega^a_{~b} \la_{Lp}^{\alpha b} , \\
   \lb_{p\:\!\dot{\alpha}}^{~\;a} & ~\to~
      \lb_{p\:\!\dot{\beta}}^{~a} (S^{\dagger})^{\dot{\beta}}_{~\dot{\alpha}}
    ~=~ \omega_{~b}^{a} \lb_{Lp\:\!\dot{\alpha}}^{~~\;b} ,  \qquad \quad
   \lb_{p}^{\dot{\alpha}a} ~\to~
      (S^{\dagger\;\!-1})^{\dot{\alpha}}_{~\dot{\beta}} \lb_{p}^{\dot{\beta}a}
    ~=~ \omega_{~b}^{a} \lb_{Lp}^{\dot{\alpha}b} ,
\label{lorentz2littlegroupmassive}
\eeal
where $\omega \in {\rm SU}(2)$ correspond to
the ${\rm SO}(3)$ rotations in the rest frame of the massive particle momentum.
These transformations are a massive analogue of \eqn{lorentz2littlegroupmassless}.
Furthermore, the momentum decomposition~\eqref{momentumbispinormassive}
implies the two-dimensional version of the Dirac equation
\beal
   p^{\dot{\alpha}\alpha} \la_{p\:\!\alpha}^{~\;a}
    = m \lb_{p}^{\dot{\alpha}a} , \qquad \quad
   p_{\alpha\dot{\alpha}} \lb_{p}^{\dot{\alpha}a} 
    = m \la_{p\:\!\alpha}^{~\;a} .
\label{diracmassive}
\eeal
For further convenience, let us rewrite the above identities in the spinor bra-ket notation:
\be
   \begin{aligned}
   \ket{p^a}_{\;\!\!\alpha}\;\![p_a|_{\dot{\beta}}& = p_{\alpha\dot{\beta}} \\
   |p^a]^{\dot{\alpha}}\:\!\bra{p_a}^{\beta}& =-p^{\dot{\alpha}\beta} \\
   \ket{p^a}_{\;\!\!\alpha} \bra{p_a}^{\beta} & =-m\:\!\delta_\alpha^\beta \\
   |p^a]^{\dot{\alpha}}\;\:\!\![p_a|_{\dot{\beta}} &
    = m\:\!\delta^{\dot{\alpha}}_{\dot{\beta}}
   \end{aligned} \qquad \quad
   \begin{aligned}
   p^{\dot{\alpha}\beta} \ket{p^a}_{\;\!\!\beta} & = m |p^a]^{\dot{\alpha}} \\
   p_{\alpha\dot{\beta}} |p^a]^{\dot{\beta}} & = m \ket{p^a}_{\;\!\!\alpha} \\
   \bra{p^a}^{\alpha} p_{\alpha\dot{\beta}} & = -m [p^a|_{\dot{\beta}} \\
   [p^a|_{\dot{\alpha}} p^{\dot{\alpha}\beta} & = -m \bra{p^a}^{\beta} 
   \end{aligned} \qquad \quad
   \begin{aligned}
   \braket{p^a p^b} & = -m\:\!\epsilon^{ab} \\
   [p^a p^b] & = m\:\!\epsilon^{ab} .
   \end{aligned}
\label{massivecompleteness}
\ee

As an explicit spinor realization, one may use~\cite{Arkani-Hamed:2017jhn}
\be
   \la_{p\:\!\alpha}^{~\;a}\!=\!
      \begin{pmatrix}       \sqrt{E\!-\!P}\:\!\cos\frac{\theta}{2} &
\!\!\!\!-\sqrt{E\!+\!P}\:\!e^{-i\varphi}\:\!\!\sin\frac{\theta}{2} \\
        \!\sqrt{E\!-\!P}\:\!e^{i\varphi}\:\!\!\sin\frac{\theta}{2} &
                            \sqrt{E\!+\!P}\:\!\cos\frac{\theta}{2}
      \end{pmatrix}\!, ~~
   \lb_{p\:\!\dot{\alpha}}^{~\;a}\!=\!
      \begin{pmatrix}
       \!-\sqrt{E\!+\!P}\:\!e^{i\varphi}\:\!\!\sin\frac{\theta}{2} &
                         \!-\sqrt{E\!-\!P}\:\!\cos\frac{\theta}{2} \\
                            \sqrt{E\!+\!P}\:\!\cos\frac{\theta}{2} &
\!\!\!\!-\sqrt{E\!-\!P}\:\!e^{-i\varphi}\:\!\!\sin\frac{\theta}{2}
      \end{pmatrix}\!,
\label{massivespinorsolutionangles}
\ee
given a massive momentum expressible as
$p^\mu=(E,P\cos\varphi \sin\theta,P\sin\varphi \sin\theta,P\cos\theta)$,
such that $E^2\!-P^2=m^2$.
A more detailed implementation is given in \app{sec:massivespinors}.

\subsection{Dirac spinors and spin}

In this paper, we wish to study massive quarks
that are traditionally described in terms of the Dirac spinors.
Hence it may be illuminating to consider
how the Weyl spinors~\eqref{momentumbispinormassive}
naturally unify into the Dirac spinors:\footnote{We use the Dirac matrices in the Weyl basis,
$\gamma^\mu=\big(\begin{smallmatrix} 0 & \sigma^{\mu} \\
                    \bar{\sigma}^{\mu} & 0 \end{smallmatrix}\big)$, hence
$\gamma^{5} = \big(\begin{smallmatrix} -1 & 0 \\ 0 & 1 \end{smallmatrix}\big)$ and
$\vec{\Sigma} = \gamma^0\vec{\gamma}\;\!\gamma^5\!=
 \big(\begin{smallmatrix}\vec{\:\!\sigma} & 0 \\ 0 & \vec{\:\!\sigma}\end{smallmatrix}\big)$.}
\begin{subequations} \begin{align}
   u_{p}^{Aa} &
    = \begin{pmatrix} \la_{p\:\!\alpha}^{~\;a} \\ \lb_{p}^{\dot{\alpha}a}
      \end{pmatrix} , \qquad \quad
   \bar{u}_{p\:\!A}^{a}
    = \begin{pmatrix} -\la_{p}^{\alpha a} \\ ~~\lb_{p\:\!\dot{\alpha}}^{~\;a}
      \end{pmatrix}
   \quad\;\Rightarrow \qquad
   \left\{
   \begin{aligned}
    & (\!\not{\!p}-m) u_{p}^{a} = \bar{u}_{p}^{a} (\!\not{\!p}-m) = 0 , \\
    & \bar{u}_{p}^{a} u_{p}^{b} = 2m\epsilon^{ab} , \\
    & \bar{u}_{p}^{a} \gamma^\mu\:\!\!u_{p}^{b} = 2p^\mu\;\!\!\epsilon^{ab} , \\
    & u_{p}^a \bar{u}_{p\:\!a} = u_{p}^{a} \epsilon_{ab} \bar{u}_{p}^{b}
    = \not{\!p}+m ,
   \end{aligned}
   \right.\!\!
\label{diracmassiveu} \\
   v_{p}^{Aa} &
    = \begin{pmatrix}-\la_{p\:\!\alpha}^{~\;a} \\ ~~\lb_{p}^{\dot{\alpha}a}
      \end{pmatrix} , \qquad\;
   \bar{v}_{p\:\!A}^{a}
    = \begin{pmatrix} \la_{p}^{\alpha a} \\ \lb_{p\:\!\dot{\alpha}}^{~\;a}
      \end{pmatrix}
   \qquad \Rightarrow \qquad
   \left\{
   \begin{aligned}
    & (\!\not{\!p}+m) v_{p}^{a} = \bar{v}_{p}^{a} (\!\not{\!p}+m) = 0 , \\
    & \bar{v}_{p}^{a} v_{p}^{b} = 2m\epsilon^{ab} , \\
    & \bar{v}_{p}^{a} \gamma^\mu\:\!\!v_{p}^{b} =-2p^\mu\;\!\!\epsilon^{ab} , \\
    & v_{p}^a \bar{v}_{p\:\!a} = v_{p}^{a} \epsilon_{ab} \bar{v}_{p}^{b}
    = -\!\!\not{\!p}+m .
   \end{aligned}
   \right.\!\!
\label{diracmassivev}
\end{align} \label{diracmassiveuv}%
\end{subequations}
This choice of $\bar{u}_{p}^{a}$ and $\bar{v}_{p}^{a}$ is consistent with
the conjugation properties $(u^a_p)^\dagger=\sgn(p^0)\bar{u}_{p\:\!a} \gamma^0$,
$(v^a_p)^\dagger=-\sgn(p^0)\bar{v}_{p\:\!a} \gamma^0$,
assuming that the constituent Weyl spinors
are parametrized as detailed in \app{sec:massivespinors}.

We can treat these spinors as quantum-mechanical wavefunctions
and compute the expectation values of the spin operator $\vec{\Sigma}/2$,
where $\Sigma^i \equiv i\epsilon^{ijk}\gamma_j\gamma_k/2$.
Given the spinor parametrization~\eqref{massivespinorsolutionangles},
we obtain the three-dimensional spin vector
\be\!\!\!
   \vec{\:\!s}\:\!(u_p^a)
    \equiv \frac{1}{2} \frac{ u_p^{a\dagger}\;\!\vec{\Sigma}\;\!u_p^{a} }
                            { u_p^{a\dagger} u_p^{a} }
    = \frac{ \bar{u}_{p\:\!a}\;\!\vec{\gamma}\,\gamma^5\;\!u_p^{a} }
           { 2\;\!\bar{u}_{p\:\!a} \gamma^0 u_p^{a} }
    = \frac{(-1)^{a-1}\!\!\!}{2} (\cos\varphi \sin\theta, \sin\varphi \sin\theta, \cos\theta)
    \equiv \frac{(-1)^{a-1}\!\!\!}{2} \hat{p} .
\label{spin3vector}
\ee
Therefore, the spinors~\eqref{massivespinorsolutionangles} have definite $\pm 1/2$ helicities,
\ie the eigenvalues of the helicity operator $h = {\hat{p}\cdot\vec{\Sigma}}/2$,
which is a conserved quantity for a one-particle state.

To delve into the subject of spin a bit further,
we rewrite the massive spinor parametrization~\eqref{massivespinorsolutionangles} as
\begin{subequations} \begin{align}
   \la_{p\:\!\alpha}^{~\;a} &\!=
      \sqrt{E\!+\!P}
      \begin{pmatrix}   \!-e^{-i\varphi}\:\!\!\sin\frac{\theta}{2} \\
                                              \cos\frac{\theta}{2}
      \end{pmatrix}_{\!\alpha}\!\!\!\otimes\!
      \begin{pmatrix} 0 \\ 1
      \end{pmatrix}^{\!a}\!
    + \frac{m}{\sqrt{E\!+\!P}}
      \begin{pmatrix}                         \cos\frac{\theta}{2} \\
                            e^{i\varphi}\:\!\!\sin\frac{\theta}{2}  
      \end{pmatrix}_{\!\alpha}\!\!\!\otimes\!
      \begin{pmatrix} 1 \\ 0
      \end{pmatrix}^{\!a}\!, \\
   \lb_{p}^{\dot{\alpha}a} &\!=
      \sqrt{E\!+\!P}
      \begin{pmatrix}                         \cos\frac{\theta}{2} \\
                            e^{i\varphi}\:\!\!\sin\frac{\theta}{2}  
      \end{pmatrix}^{\!\dot{\alpha}}\!\!\!\otimes\!
      \begin{pmatrix} 1 \\ 0
      \end{pmatrix}^{\!a}\!
    + \frac{m}{\sqrt{E\!+\!P}}
      \begin{pmatrix}   \!-e^{-i\varphi}\:\!\!\sin\frac{\theta}{2} \\
                                              \cos\frac{\theta}{2}
      \end{pmatrix}^{\!\dot{\alpha}}\!\!\!\otimes\!
      \begin{pmatrix} 0 \\ 1
      \end{pmatrix}^{\!a}\!,
\end{align} \label{massivespinorsolutiondecomposition}%
\end{subequations}
which makes obvious
the smooth limit of the massive spinors~$\la_{p\:\!\alpha}$ and~$\lb_{p}^{\dot{\alpha}}$
to their massless homonymes~\eqref{spinorsolutionangles}:
\be
   \la_{p\:\!\alpha}^{~\;a} \xrightarrow[m\to0]{}
      \la_{p\:\!\alpha} \zeta^a_- , \qquad \quad
   \lb_{p}^{\dot{\alpha}a} \xrightarrow[m\to0]{}
      \lb_{p}^{\dot{\alpha}}\:\!\zeta^a_+ ,
\ee
where $\zeta^a_-\equiv(0,1)$ and $\zeta^a_+\equiv(1,0)$.
To rephrase this in a more general way,
we can introduce two-dimensional spinors
$\la_{p\:\!\alpha}$ and $\eta_{p\:\!\alpha}$ such that
$\la_{p\:\!\alpha}^{~\;a}$ and $\lb_{p}^{\dot{\alpha}a}$ decompose as
\be
   \la_{p\:\!\alpha}^{~\;a} =
      \la_{p\:\!\alpha} \zeta^a_-\!
    + \eta_{p\:\!\alpha} \zeta^a_+ , \qquad \quad
   \lb_{p}^{\dot{\alpha}a} =
      \lb_{p}^{\dot{\alpha}}\:\!\zeta^a_+\!
    - \tilde{\eta}_{p}^{\dot{\alpha}}\:\!\zeta^a_- , \qquad \quad
   \braket{\la_p \eta_p} = [\tilde{\eta}_p \lb_p] = m .
\label{massivespinorsolutiondecomposition2}
\ee
The massive momentum is now expressed as a sum of two null momenta:
\be
   p_{\alpha\dot{\alpha}} = \la_{p\:\!\alpha} \lb_{p\:\!\dot{\alpha}}
    + \eta_{p\:\!\alpha} \tilde{\eta}_{p\:\!\dot{\alpha}} ,
\ee
which gives a link to the massive extension of the massless spinor-helicity formalism
used previously in the literature~\cite{Kleiss:1986qc,Dittmaier:1998nn,Schwinn:2005pi}.
We make this link precise in \sec{sec:checks} below.

Now let us discuss a subtle point concerning spin.
Traditional quantum-mechanical spin operators
are thought of as acting on the ${\rm SU}(2)$ indices,
which seem to correspond to the little group.
The spin of the decomposition~\eqref{massivespinorsolutionangles}
points along the three-momentum $\vec{\:\!p}$,
whereas the little-group vectors $\zeta^a_{\pm}$ describe states
with spin direction along the $z$-axis.
In other words, the massive Weyl spinors~\eqref{massivespinorsolutionangles}
convert the physical helicity operator
$h = {\hat{p}\cdot\vec{\Sigma}}/2 =
 \Big(\begin{smallmatrix}
         \hat{p}\cdot\vec{\:\!\sigma}\! & 0 \\
         0 & \hat{p}\cdot\vec{\:\!\sigma}\:\!
      \end{smallmatrix}\Big)/2$
to $\sigma^3/2$:
\be
   (\hat{p}\cdot\vec{\:\!\sigma})_{\alpha}^{~\beta} \la_{p\:\!\beta}^{~a}
    = \sigma^{3\:\!a}_{~~\,b} \la_{p\:\!\alpha}^{~b} , \qquad \quad
   (\hat{p}\cdot\vec{\:\!\sigma})^{\dot{\alpha}}_{~\dot{\beta}} \lb_{p}^{\dot{\beta}a}
    = \sigma^{3\:\!a}_{~~\,b} \lb_{p}^{\dot{\alpha}b} .
\label{helicity2spinz}
\ee
This should be regarded as a nice feature
of the parametrization~\eqref{massivespinorsolutionangles}
rather than an inconsistency.
Indeed, the little-group ${\rm SU}(2)$ transformations
correspond to ${\rm SO}(3)$ rotations in the rest frame of the massive particle,
in which $p^\mu_{\text{rest}} = (m,\vec{\:\!0})$,
whereas the spinorial matrices
$\vec{\:\!\sigma}_{\alpha}^{~\beta}=\vec{\:\!\sigma}^{\dot{\alpha}}_{~\dot{\beta}}$
generate rotations in the boosted frame where $p^\mu=(E,\vec{\:\!p})$.
It is therefore convenient that the spinorial $(\hat{p}\cdot\vec{\:\!\sigma})$,\footnote{In fact, the other two spatial directions corresponding to the little-group matrices
$\sigma^{1\:\!a}_{~~\,b}$ and $\sigma^{2\:\!a}_{~~\,b}$
in the sense of \eqn{helicity2spinz} turn out to be complex for any nonzero $\vec{\:\!p}$.
The corresponding spin-projection operators are thus not hermitian,
and there is no unitary intertwining operator
between the two representations of the complete spin operator $\vec{\:\!\sigma}/2$.
Indeed, such an operator would have to involve a boost transformation to the rest frame,
which lies outside the rotational ${\rm SU}(2)$.}
taken along the momentum direction,
are converted to the simplest of the Pauli matrices, $\sigma^{3\:\!a}_{~~\,b}$.

In principle, one can easily break the above property
by ${\rm SU}(2)$-rotating the spin states.
Apart from losing
the relatively simple parametrization~\eqref{massivespinorsolutionangles},
this would mix the pure helicity eigenstates
and produce wavefunctions with a spin quantization axis other than the momentum,
and therefore undetermined helicity.
The massive spinor-helicity formalism of \rcite{Arkani-Hamed:2017jhn} reviewed here
allows to easily switch that axis, and this is precisely what we do in \sec{sec:checks}
in order to compare our results with the literature.

\section{Four-point amplitudes}
\label{sec:4pt}

In this section, we demonstrate the use of the various spinors discussed above
by dissecting one full color-dressed amplitude.
It is convenient to consider the simple case of
one massive quark-antiquark pair and two gluons of opposite helicity.
Their scattering amplitude has three Feynman diagrams:\footnote{We normalize the group generators to obey
$\Tr(T^{a} T^{b})=\delta^{ab}$ and $[T^a,T^b]=\tf^{abc}T^c$
and regard all particle momenta as outgoing.
We use slashed matrices $\not{\!p}$ to denote either
$\gamma^\mu p_\mu$, $\sigma^\mu p_\mu$ or $\bar{\sigma}^\mu p_\mu$,
depending on the spinors surrounding them.
In expressions like $\bra{i}j|k]\equiv\bra{i}\!\not{\!p}_j|k]=\bra{i}p_j|k]$
the slash can be omitted.}
\begin{subequations} \begin{align}\!\!\!
   \scalegraph{0.9}{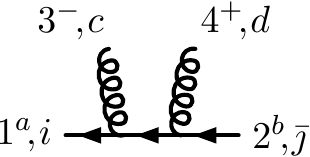}\!\!\!
      & =-\frac{i}{2}\,\frac{T_{i \bar k}^c T_{k \bar\jmath}^d}{s_{13}\!-\!m^2}\,
        ( \bar{u}_1^a\!\!\not{\;\!\!\varepsilon}_3^- (\!\not{\!p}_{13}\!+\!m)
                     \!\!\not{\;\!\!\varepsilon}_4^+ v_2^b )
        \equiv \frac{c_1 n_1}{D_1} ,
\label{QQgg1} \\\!\!\!
   \scalegraph{0.9}{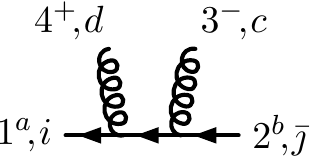}\!\!\!
      & =-\frac{i}{2}\,\frac{T_{i \bar k}^d T_{k \bar\jmath}^c}{s_{14}\!-\!m^2}\,
        ( \bar{u}_1^a\!\!\not{\;\!\!\varepsilon}_4^+ (\!\not{\!p}_{14}\!+\!m)
                     \!\!\not{\;\!\!\varepsilon}_3^- v_2^b )
        \equiv \frac{c_2 n_2}{D_2} ,
\label{QQgg2} \\\!\!\!
   \scalegraph{0.9}{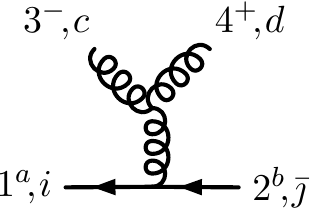}\!\!\! &
      \begin{aligned}\;\!=
          \frac{i}{2}\frac{\tilde{f}^{cde} T_{i \bar\jmath}^e}{s_{34}}\,
          \Big\{ (\varepsilon_3^-\!\!\cdot\!\varepsilon_4^+)
                 ( \bar{u}_1^a (\!\not{\!p}_3-\!\not{\!p}_4) v_2^b )
                + 2 (p_{4}\!\cdot\!\varepsilon_3^-)
                  ( \bar{u}_1^a\!\!\not{\;\!\!\varepsilon}_4^+ v_2^b ) & \\
               -\,2 (p_{3}\!\cdot\!\varepsilon_4^+)
                  ( \bar{u}_1^a\!\!\not{\;\!\!\varepsilon}_3^- v_2^b ) & \Big\}
        \equiv \frac{c_3 n_3}{D_3} .
      \end{aligned}\!\!\!
\label{QQgg3}
\end{align} \label{QQgg}%
\end{subequations}
Now let us recast the above numerators in the spinor-helicity formalism
by plugging in the Dirac spinors~\eqref{diracmassiveuv}
and the polarization vectors~\eqref{polvectors},
\begin{subequations} \begin{align}
   n_1 = \frac{-i}{[3\;\!q_3]\braket{4\;\!q_4}}
      \Big\{ \braket{1^a 3} [q_3|p_{13}\ket{q_4} [4\;\!2^b]
           + [1^a q_3] \bra{3}p_{13}|4] \braket{q_4 2^b} & \\
          -\,m \braket{1^a 3} [q_3 4] \braket{q_4 2^b}
           - m [1^a q_3] \braket{3\;\!q_4} [4\;\!2^b] & \Big\} , \nn \\
   n_2 = \frac{-i}{[3\;\!q_3]\braket{4\;\!q_4}}
      \Big\{ \braket{1^a q_4} [4|p_{14}\ket{3} [q_3 2^b]
           + [1^a 4] \bra{q_4}p_{14}|q_3] \braket{3\;\!2^b} & \\
          -\,m \braket{1^a q_4} [4\;\!q_3] \braket{3\;\!2^b}
           - m [1^a 4] \braket{q_4 3} [q_3 2^b] & \Big\} , \nn \\
   n_3 = \frac{-i}{[3\;\!q_3]\braket{4\;\!q_4}}
      \bigg\{\!\!-\!\frac{1}{2} \braket{3\;\!q_4} [4\;\!q_3]
             \big( \bra{1^a}p_3\!-\!p_4|2^b]
                 + [1^a|p_3\!-\!p_4\ket{2^b} \big) \!\!\!\!\!\!\!\!\!\; & \\
           -\,\bra{3}4|q_3]
             \big( \braket{1^a q_4} [4\;\!2^b] + [1^a 4] \braket{q_4 2^b} \big)
           + \bra{q_4}3|4]  \!\!\!\!\!\!\!\!\!\!\;&\,\,\,\,\,\,\,\:
             \big( \braket{1^a 3} [q_3 2^b] + [1^a q_3] \braket{3\;\!2^b} \big)\!
      \bigg\} ,\! \nn
\end{align} \label{QQggSH}%
\end{subequations}
where for brevity we label spinors as $\ket{i}\equiv\ket{p_i}$, \etc
The numerators~\eqref{QQggSH} may seem complicated, which is due to their
explicit gauge dependence on the gluonic reference vectors~$q_3$ and~$q_4$.
Incidentally, one can check that for any such gauge choice
they nontrivially satisfy the kinematic-algebra relation $n_1-n_2=n_3$,
which is color-dual to the commutation relation $c_1-c_2=c_3$
\cite{Johansson:2014zca,Johansson:2015oia}.
A very beneficial gauge choice is $q_3=p_4$ and $q_4=p_3$, for which
\be
   n_1 = n_2 = \frac{i}{s_{34}} \bra{3}1|4]
      \big( \braket{1^a 3} [2^b 4] + [1^a 4] \braket{2^b 3} \big) ,
   \qquad \quad n_3 = 0 .
\ee
We can thus write simple closed-form expressions for all three color-ordered amplitudes
\begin{subequations} \begin{align}
\label{QQggA1}
   A(\u{1}^a\:\!\!,\o{2}^b\:\!\!,3^-\!,4^+) & \equiv \frac{n_2}{D_2} - \frac{n_3}{D_3}
    = \frac{i \bra{3}1|4]}{(s_{14}\!-\!m^2)s_{34}}
      \big( \braket{1^a 3} [2^b 4] + [1^a 4] \braket{2^b 3} \big) , \\*
\label{QQggA2}
   A(\u{1}^a\:\!\!,\o{2}^b\:\!\!,4^+\!,3^-) & \equiv \frac{n_1}{D_1} + \frac{n_3}{D_3}
    = \frac{i \bra{3}1|4]}{(s_{13}\!-\!m^2)s_{34}}
      \big( \braket{1^a 3} [2^b 4] + [1^a 4] \braket{2^b 3} \big) , \\*
\label{QQggA3}
   A(\u{1}^a\:\!\!,3^-\!,\o{2}^b\:\!\!,4^+) & \equiv-\frac{n_1}{D_1} - \frac{n_2}{D_2}
    = \frac{i \bra{3}1|4]}{(s_{13}\!-\!m^2)(s_{14}\!-\!m^2)}
      \big( \braket{1^a 3} [2^b 4] + [1^a 4] \braket{2^b 3} \big) .\!
\end{align} \label{QQggA}%
\end{subequations}
These evidently obey
the Kleiss-Kuijf relation $A_{1243}+A_{1234}+A_{1324}=0$ \cite{Kleiss:1988ne},
as well as the Bern-Carrasco-Johansson (BCJ)
relation~\cite{Bern:2008qj,Johansson:2015oia,delaCruz:2015dpa}
\be
   (s_{14}-m^2) A(\u{1}^a\:\!\!,\o{2}^b\:\!\!,3^-\!,4^+)
 = (s_{13}-m^2) A(\u{1}^a\:\!\!,\o{2}^b\:\!\!,4^+\!,3^-) .
\label{bcj4pt}
\ee
The full color-dressed amplitude can thus be constructed
from a single linearly independent color-ordered amplitude as~\cite{DelDuca:1999rs,Johansson:2015oia}
\be
   {\cal A}(1^a_{i}\:\!\!,2^b_{\bar \jmath},3^-_c\!,4^+_d)
    = i\bigg[ \frac{T_{i \bar k}^c T_{k \bar\jmath}^d}{(s_{13}\!-\!m^2)s_{34}}
            + \frac{T_{i \bar k}^d T_{k \bar\jmath}^c}{(s_{14}\!-\!m^2)s_{34}}
       \bigg] \bra{3}1|4]
       \big( \braket{1^a 3} [2^b 4] + [1^a 4] \braket{2^b 3} \big) .
\ee
It is interesting to note~\cite{Bjerrum-Bohr:2013bxa}
that the gluonic color-ordered amplitude~\eqref{QQggA3}
is also the correct QED amplitude~\cite{Arkani-Hamed:2017jhn}
(up to a factor of $-2$ due to the color-generator conventions).

Note that the above amplitudes are gauge-invariant and could have been reduced
from the numerators~\eqref{QQggSH} to the expressions~\eqref{QQggA}
for any choice of reference vectors~$q_3$ and~$q_4$.
This illustrates why in general, at least in analytic calculations,
it is better to avoid dealing with gauge-dependent objects
and compute gauge-invariant quantities directly.
Such a way to derive the above amplitudes would be
via the BCFW on-shell recursion~\cite{Britto:2004ap,Britto:2005fq}
starting from the three-point amplitudes
\be\!\!\!\!\!\!
   \left.
   \begin{aligned}
   {\cal A}(1^a_{i}\:\!\!,2^b_{\bar \jmath},3^+_c) &
    =-\frac{i T_{i \bar\jmath}^c}{\braket{3\;\!q}}
      \big( \braket{1^a q}[2^b 3] + [1^a 3]\braket{2^b q} \big)
    =-i T_{i \bar\jmath}^c \frac{\braket{1^a 2^b} [3|1\ket{q}}{m\braket{3\;\!q}} \\
   {\cal A}(1^a_{i}\:\!\!,2^b_{\bar \jmath},3^-_c) &
    = \frac{i T_{i \bar\jmath}^c}{[3\;\!q]}
      \big( \braket{1^a 3}[2^b q] + [1^a q]\braket{2^b 3} \big)
    = i T_{i \bar\jmath}^c \frac{[1^a 2^b] \bra{3}1|q]}{m[3\;\!q]}
   \end{aligned}
   \right\} =\!\!\scalegraph{1.0}{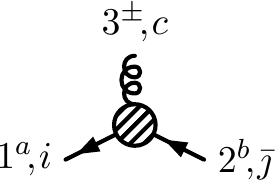}\!\!\!\!\!\!\!\!\!\!\!
\label{QQgA}%
\ee
These make sense on complex on-shell kinematics
and are independent of the gauge-boson reference vector $q$,
despite not looking that way (this feature is explained in \rcite{Arkani-Hamed:2017jhn}).

\begin{figure}[t]
\centering
\vspace{-7pt}
$ \scalegraph{1.0}{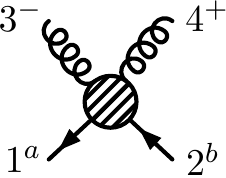}~=~\scalegraph{1.0}{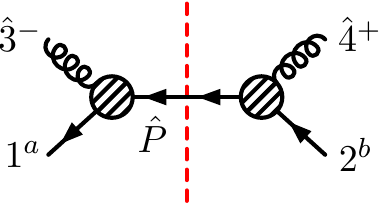} $
\vspace{-5pt}
\caption{\small Graphic representation for the BCFW derivation~\eqref{bcfwQQgg}
of $A(\u{1}^a\:\!\!,\hat{3}^-\!,\hat{4}^+\!,\o{2}^b)$}
\label{fig:bcfwQQgg}
\end{figure}

To reduce the four-point amplitude to the three-point ones,
we apply a simple massless-spinor shift
\be
   |\hat{3}] \equiv |3] - z |4] , \qquad \quad
   \ket{\hat{4}} \equiv \ket{4} + z \ket{3} ,
\label{BCFWshift34}
\ee
which preserves momentum conservation and the on-shell conditions for any complex $z$.
Cauchy's integral theorem then localizes the four-point amplitude
$A(\u{1}^a\:\!\!,\o{2}^b\:\!\!,4^+\!,3^-)$ on the only pole
$z_{13}=\bra{3}1|3]/\bra{3}1|4]$ corresponding
to the quark propagator $\hat{P} \equiv p_1+\hat{p}_3$, see \fig{fig:bcfwQQgg}.
Hence the amplitude factorizes as a product of two three-point amplitudes
with complex momenta:
\begin{align}
   A(\u{1}^a\:\!\!,\o{2}^b\:\!\!,&\,4^+\!,3^-)
    = \Res_{z=z_{13}\!\!} A(\u{1}^a\:\!\!,\hat{3}^-\!,\hat{4}^+\!,\o{2}^b)
    = A(\u{1}^a\:\!\!,\hat{3}^-\!,\!-\hat{P}^c) \frac{i}{s_{13}-m^2}
      A(\hat{P}_c,\hat{4}^+\!,\o{2}^b) \nn \\* &
    = \frac{i}{(s_{13}\!-\!m^2)[\hat{3}\;\!q_3]\braket{\hat{4}\;\!q_4}}
      \big( \braket{1^a 3}[-\hat{P}^c|q_3] + [1^a q_3]\braket{-\hat{P}^c|3} \big)
      \big( \braket{\hat{P}_c\;\!q_4}[2^b 4]
          + [\hat{P}_c\;\!4]\braket{2^b q_4} \big)\nn \\* &
    = \frac{-i}{(s_{13}\!-\!m^2)[3\;\!4]\braket{4\;\!3}}
      \big( \braket{1^a 3}[4\hat{P}^c] - [1^a 4]\braket{3\hat{P}^c} \big)
      \big( \braket{\hat{P}_c\;\!3}[2^b 4]
          + [\hat{P}_c\;\!4]\braket{2^b 3} \big) \nn \\* &
\label{bcfwQQgg}
    = \frac{-i}{(s_{13}\!-\!m^2) s_{34}}
      \Big\{ \braket{1^a 3} [2^b 4] [4\hat{P}^c] \braket{\hat{P}_c\;\!3}
           + \braket{1^a 3} \braket{2^b 3} [4\hat{P}^c] [\hat{P}_c\;\!4] \\* &
             \qquad \qquad \qquad~~
           - [1^a 4] [2^b 4] \braket{3\hat{P}^c} \braket{\hat{P}_c\;\!3}
           - [1^a 4] \braket{2^b 3} \braket{3\hat{P}^c} [\hat{P}_c\;\!4]
      \Big\} \nn \\* &
    = \frac{i\bra{3}1|4]}{(s_{13}\!-\!m^2) s_{34}}
      \big( \braket{1^a 3} [2^b 4] + [1^a 4] \braket{2^b 3} \big) . \nn
\end{align}
Here we chose the reference vectors as $q_3=p_4$, $q_4=p_3$ to remove most $z$-dependence
as early as possible.
Otherwise, the spinor products of $q_3$ and $q_4$ would cancel anyway,
but only after plugging in the specific on-shell solutions for
$|\hat{3}]$, $\ket{\hat{4}}$ and $\hat{P}^\mu$ and using various Schouten identities.
In the last transition of \eqn{bcfwQQgg},
we also reduced the sum over the spin label $c$ of the intermediate quark
using the completeness relations~\eqref{massivecompleteness}.

As a simple check, we verify that the massless limit
corresponds to the well-known Parke-Taylor MHV amplitudes~\cite{Parke:1986gb}:
\be\!\!
   {\cal A}(1^a_{i}\:\!\!,2^b_{\bar \jmath},3^-_c\!,4^+_d) \xrightarrow[m\to0]{}
      i \bigg[ \frac{ T_{i \bar k}^c T_{k \bar\jmath}^d }
                    { \braket{1\;\!2}\braket{2\;\!4}\braket{4\;\!3}\braket{3\;\!1} }
             + \frac{ T_{i \bar k}^d T_{k \bar\jmath}^c }
                    { \braket{1\;\!2}\braket{2\;\!3}\braket{3\;\!4}\braket{4\;\!1} }
        \bigg]\!
      \begin{pmatrix} ~\,0 \!\!&\!\! \braket{1\;\!3} \braket{2\;\!3}^3 \\
                      -\braket{1\;\!3}^3 \braket{2\;\!3} \!\!&\!\! 0 \end{pmatrix} .
\ee

It is even simpler calculation, either with the Feynman diagrams
or via the on-shell recursion,
to find the amplitudes with two quarks and two positive-helicity gluons
\be\!\!\!
   A_{1234}
    = \frac{i\;\!m \braket{1^a 2^b} [3\;\!4]}
           {(s_{14}\!-\!m^2) \braket{3\;\!4}} , \quad~
   A_{1243}
    = \frac{i\;\!m \braket{1^a 2^b} [3\;\!4]}
           {(s_{13}\!-\!m^2) \braket{3\;\!4}}  , \quad~
   A_{1324}
    = \frac{\!\!\!\!\;-i\;\!m \braket{1^a 2^b} [3\;\!4]^2}
           {(s_{13}\!-\!m^2)(s_{14}\!-\!m^2)} .\!
\label{QQggAP}%
\ee

\section{All-multiplicity amplitudes}
\label{sec:npt}

In this section we turn to the main calculations of this note ---
two infinite families of color-ordered amplitudes with one massive quark-antiquark pair.

\subsection{All-plus amplitudes with two quarks}
\label{sec:allplus}

\begin{figure}[t]
\centering
\vspace{-7pt}
$ \scalegraph{1.0}{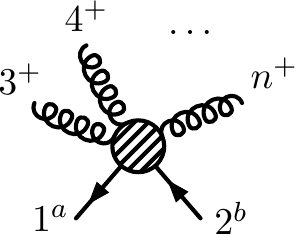}~=~\scalegraph{1.0}{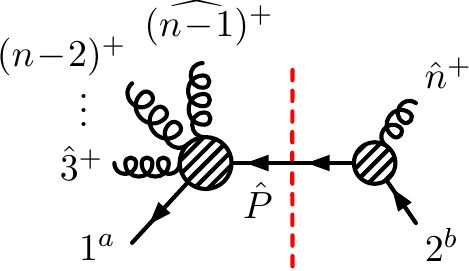}~+~~\scalegraph{1.0}{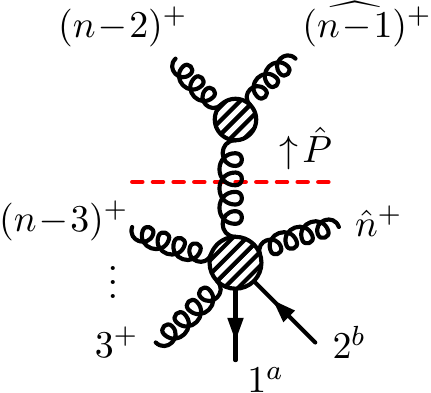} $
\vspace{-5pt}
\caption{\small Graphic representation of the BCFW recursion step~\eqref{bcfwAP}
for $A(\u{1}^a\:\!\!,3^+\!,4^+\!,\dots,n^+\!,\o{2}^b)$}
\label{fig:bcfwAP}
\end{figure}

The $n$-point amplitude for a quark-antiquark pair and $n-2$ positive-helicity gluons equals
\be\!\!
   A(\u{1}^a\:\!\!,3^+\!,4^+\!,\dots,n^+\!,\o{2}^b)
    = \frac{i\;\!m \braket{1^a 2^b}
             [3|\prod_{j=3}^{n-2}\!\big\{\!\!\not{\!p}_{13 \dots j}\!\not{\!p}_{j+1}
                                        +(s_{13 \dots j}-m^2) \big\}|n] }
           { (s_{13}\!-\!m^2)(s_{134}\!-\!m^2)\dots(s_{13\dots(n-1)}\!-\!m^2)\,
             \braket{34}\braket{45}\dots\braket{n\!-\!1|n} } .\!\!
\label{QQggnAP}
\ee
It is easiest derived using the BCFW recursion~\cite{Britto:2004ap,Britto:2005fq}.
To set up the induction, we check that for $n=4$ the formula~\eqref{QQggnAP} visibly
reduces to the four-point amplitude $A_{1243}$ in \eqn{QQggAP}.
For the inductive step, we choose to shift the gluonic spinors
\be
   |\widehat{n\!-\!1}] \equiv |n\!-\!1] - z |n] , \qquad \quad
   \ket{\hat{n}} \equiv \ket{n} + z \ket{n\!-\!1} .
\label{BCFWshiftn1n}
\ee
Then well-known arguments~\cite{Schwinn:2007ee,ArkaniHamed:2008yf,Britto:2012qi}
guarantee a vanishing boundary behavior at $z\to\infty$.
There are two potential contributions in the on-shell recursion:
\beal
   A(\u{1}^a\:\!\!,3^+\!,4^+\!,\dots,n^+\!,\o{2}^b)
    = A(\u{1}^a\:\!\!,3^+\!,\dots,(n\!-\!2)^+\!,(\widehat{n\!-\!1})^+\!,\!-\hat{P}^c)
      \frac{i}{s_{2n}-m^2} A(\hat{P}_c,\hat{n}^+\!,\o{2}^b) & \\
    + A((n\!-\!2)^+\!,(\widehat{n\!-\!1})^+\!,\!-\hat{P}^-)
      \frac{i}{s_{(n-2)(n-1)}}
      A(\u{1}^a\:\!\!,3^+\!,\dots,(n\!-\!3)^+\!,\hat{P}^+\!,\hat{n}^+\!,\o{2}^b) & .\!
\label{bcfwAP}
\eeal
Any other $z$-dependent propagator would factorize
on a vanishing purely gluonic amplitude with all positive helicities.
The second pole in \eqn{bcfwAP} is localized on
\be
   \hat{s}_{(n-2)(n-1)} = 0 \qquad \Rightarrow \qquad
   z=\frac{[n\!-\!2|n\!-\!1]}{[n\!-\!2|n]} \qquad \Rightarrow \qquad
   |\widehat{n\!-\!1}] = |n\!-\!2] \frac{[n\!-\!1|n]}{[n\!-\!2|n]} ,
\ee
which sets the three-point $\overline{\text{MHV}}$ amplitude to zero.
We then compute the remaining contribution
on $z_{2n}=-{\bra{n}2|n]}/{\bra{n\!-\!1}2|n]}$:
\begin{align}
   A(\u{1}^a\:\!\!,3^+\!,\dots,n^+\!,\o{2}^b) &
    = \frac{ i \braket{1^a|\!-\!\hat{P}^c}
             [3|\prod_{j=3}^{n-3}\!\big\{\!\!\not{\!p}_{13 \dots j}\!\not{\!p}_{j+1}
                                        +(s_{13 \dots j}-m^2) \big\}|\widehat{n\!-\!1}]
             \braket{\hat{P}_c\;\!2^b} [n|\hat{P}\ket{q} }
           { (s_{13}-m^2)\dots(s_{13\dots(n-2)}-m^2)(s_{2n}-m^2)\,
             \braket{34}\dots\braket{n\!-\!2|n\!-\!1} \braket{\hat{n}\;\!q} } \nn \\ &
    = \frac{ i\;\!m \braket{1^a 2^b}
             [3|\prod_{j=3}^{n-3}\!\big\{\!\!\not{\!p}_{13 \dots j}\!\not{\!p}_{j+1}
                                        +(s_{13 \dots j}-m^2) \big\}|\widehat{n\!-\!1}]
             \bra{n\!-\!1}2|n] }
           { (s_{13}-m^2)\dots(s_{13\dots(n-1)}-m^2)\,
             \braket{34}\dots\braket{n\!-\!2|n\!-\!1} \braket{n\!-\!1|n} } ,
\end{align}
where we picked $q=p_{n-1}$ to remove most of the $z$-dependence.
To finish the proof of the closed-form expression~\eqref{QQggnAP},
it now suffices to notice that
\be
   |\widehat{n\!-\!1}] \bra{n\!-\!1}2|n] =\,\not{\!p}_{(n-1)n}\!\not{\!p}_{2}|n]
    = \big\{\!\!\not{\!p}_{13\dots(n-2)}\!\not{\!p}_{n-1}+(s_{13\dots(n-2)}-m^2)
      \big\}|n] .
\ee

It is effortless to also write the amplitude for the gluons of negative helicity:
we simply exchange the angle and square brackets as in
$\braket{p\;\!q} \leftrightarrow [q\;\!p]$ to obtain
\be\!\!\!
   A(\u{1}^a\:\!\!,3^-\!,4^-\!,\dots,n^-\!,\o{2}^b)
   =\!\frac{(-1)^n i\;\!m [1^a 2^b]
             \bra{3}\prod_{j=3}^{n-2}\!\big\{\!\!\not{\!p}_{13 \dots j}\!\not{\!p}_{j+1}
                                        +(s_{13 \dots j}-m^2) \big\}\ket{n} }
           { (s_{13}\!-\!m^2)(s_{134}\!-\!m^2)\dots(s_{13\dots(n-1)}\!-\!m^2)\,
             [34][45]\dots[n\!-\!1|n] } .\!\!
\label{QQggnAM}
\ee

\subsection{One-minus amplitudes with two quarks}
\label{sec:1minus}

In this section we again use the on-shell recursion
to derive an all-multiplicity expression
\beal
   A(\u{1}^a\:\!\!,3^-\!,4^+\!,\dots,n^+\!,\o{2}^b)
    =-\frac{ i \braket{3|1|2|3}\,
             \big( \braket{1^a 3} [2^b|1\!+\!2\ket{3}
                 + \braket{2^b 3} [1^a|1\!+\!2\ket{3} \big) }
           { s_{12} \braket{3\;\!4} \dots \braket{n\!-\!1|n}
             \braket{3|1|1\!+\!2|n} } & \\
    + \sum_{k=4}^{n-1}
      \frac{ i\;\!m \braket{3|\!\!\not{\!p}_{1}\!\not{\!p}_{3 \dots k}|3}\,
             \big( \braket{1^a 2^b}
                   \braket{3|\!\!\not{\!p}_{1}\!\not{\!p}_{3 \dots k}|3}
                 + \braket{1^a 3} \braket{2^b 3} s_{3 \dots k} \big) }
           { s_{3 \dots k}\,(s_{13\dots k}\!-\!m^2)\dots(s_{13\dots(n-1)}\!-\!m^2)\,
             \braket{3\;\!4} \dots \braket{k\!-\!1|k}
             \braket{3|\!\!\not{\!p}_{1}\!\not{\!p}_{3 \dots k}|k} } & \\ \times
      \frac{ \bra{3}\!\!\not{\!p}_{3 \dots k}
             \prod_{j=k}^{n-2}\!\big\{\!\!\not{\!p}_{13 \dots j}\!\not{\!p}_{j+1}
                                                +(s_{13 \dots j}-m^2) \big\}|n] }
           { \braket{3|\!\!\not{\!p}_{1}\!\not{\!p}_{3 \dots k}|k\!+\!1}
             \braket{k\!+\!1|k\!+\!2}\dots\braket{n\!-\!1|n} } &
\label{QQggnOM}
\eeal
for the amplitude with two quarks and $n-2$ gluons.
Here we assume negative helicity of the gluon~$3$ color-adjacent to the quark~$\u{1}$,
with all other gluon helicities positive, while the quark helicities are still left arbitrary.
In principle, this color-ordered amplitude is enough to reconstruct
the full color-dressed one-minus amplitude via the BCJ relations~\cite{Bern:2008qj,Johansson:2015oia}, such as the four-point one in \eqn{bcj4pt}.
Indeed, the BCJ relations allow to fix the position of any gluon
to be color-adjacent to the quark,
with permutations acting on the remaining gluons,
hence one may choose to fix the position of the minus-helicity gluon.
Moreover, one-plus amplitudes can also be retrieved from \eqn{QQggnOM}
via the conjugation rule $\braket{p\;\!q} \leftrightarrow [q\;\!p]$.

To prove the above formula,
we use the same ``$[3\;\!4\rangle$'' shift as in \eqn{BCFWshift34},
which gives only two non-vanishing contributions
at each step of the recursion,\footnote{Our present chiral conventions
$\la_{-p} = -\la_{p}$, $\lb_{-p} = \lb_{p}$, reviewed in \app{sec:masslessspinors},
imply $\varepsilon_{-p}^{\pm} = -\varepsilon_{p}^{\pm}$.
Therefore, crossing each gluon in an amplitude
creates an additional minus sign,
\be
   {\cal A}(p_1^\text{in},p_2^\text{out}\!,\dots,p_n^\text{out})
    = -{\cal A}(-p_1^\text{out}\!,p_2^\text{out}\!,\dots,p_n^\text{out}) .
\ee
Taking into account the completeness relation
$ \varepsilon_{p+}^{\mu} \varepsilon_{p-}^{\nu} +
  \varepsilon_{p-}^{\mu} \varepsilon_{p+}^{\nu} =
 -\eta^{\mu\nu}\!+ (p^{\mu} q^{\nu}\!+q^{\mu}p^{\nu})/(p\!\cdot\!q) $,
we conclude that gluonic poles are accounted for by the intermediate propagator
factor $-i/p^2$.
}
\begin{subequations} \begin{align}
\label{bcfwOM13}
   A(\u{1}^a\:\!\!,3^-\!,4^+\!,\dots,n^+\!,\o{2}^b)
    = A(\u{1}^a\:\!\!,\hat{3}^-\!,-\hat{P}_c)
      \frac{i}{s_{13}-m^2} A(\hat{P}^c\:\!\!,\hat{4}^+,5^+\!,\dots,\hat{n}^+\!,\o{2}^b) & \\
   +\,A(\u{1}^a\:\!\!,\hat{3}^-\!,\hat{P}^+\!,6^+\!,\dots,n^+\!,\o{2}^b)
      \frac{-i}{s_{45}} A(-\hat{P}^-\!,\hat{4}^+\!,5^+) & .
\label{bcfwOM45}
\end{align} \label{bcfwOM}%
\end{subequations}
The two residues are evaluated on the following pole kinematics:\!
\begin{subequations} \begin{align}
\label{BCFWshift34kinem13}\!\!\!
   z_{13} & = \frac{\bra{3}1|3]}{\bra{3}1|4]}\!: \quad~\,
   |\hat{3}] =\!-\frac{|1\ket{3}[34]}{\bra{3}1|4]} , \quad~\,
   \ket{\hat{4}}\!= \frac{|3\!+\!4|1\ket{3}}{\bra{3}1|4]} , \quad~\,
   \hat{P}^\mu\!= p_1^\mu - \frac{\bra{3}\sigma^\mu|1|3|4]}{2\bra{3}1|4]} ;\!\! \\ \!
\label{BCFWshift34kinem45}\!\!\!
   z_{45} & =\!-\frac{\braket{45}}{\braket{35}}\!: \quad~\,\,\,
   |\hat{3}] = \frac{|3\!+\!4\ket{5}}{\braket{35}} , \quad~~~\;
   \ket{\hat{4}}\!= \ket{5}\frac{\braket{34}}{\braket{35}} , \quad~\,
   \ket{\hat{P}}\!= \ket{5} , \quad
   |\hat{P}] = \frac{|4\!+\!5\ket{3}}{\braket{53}} .\!\!
\end{align} \label{BCFWshift34kinem}%
\end{subequations}
The residue at $z_{13}$ is computed immediately
for any $n$ using the all-plus expression~\eqref{QQggnAP},
\begin{align}
\label{BCFWshift34z13}
 & A(\u{1}^a\:\!\!,\hat{3}^-\!,-\hat{P}^c) \frac{i}{s_{13}-m^2}
      A(\hat{P}_c,\hat{4}^+,5^+\!,\dots,\hat{n}^+\!,\o{2}^b) \nn \\ &
    =-\frac{ i [1^a|\!-\!\hat{P}^c] \braket{\hat{P}_c\;\!2^b} \bra{3}1|q]\,
             [4|\prod_{j=4}^{n-2}\!\big\{\hat{\not{\!p}}_{P4 \dots j}\!\not{\!p}_{j+1}
                                        +(\hat{s}_{P4 \dots j}-m^2) \big\}|n] }
           { (s_{13}\!-\!m^2)(\hat{s}_{P4}\!-\!m^2)\dots(\hat{s}_{P4\dots(n-1)}\!-\!m^2)\,
             [\hat{3}\;\!q] \braket{\hat{4}\;\!5}\braket{5\;\!6}\dots\braket{n\!-\!1|n} } \\ &
    =-\frac{ i\;\!m \bra{3}1|4]\,
             [4|\prod_{j=4}^{n-2}\!\big\{\!\!\not{\!p}_{13 \dots j}\!\not{\!p}_{j+1}
                                        +(s_{13 \dots j}-m^2) \big\}|n]\,
             \big( \braket{1^a 2^b} \bra{3}1|4]
                 + \braket{1^a 3} \braket{2^b 3} [3\;\!4] \big) }
           { (s_{13}\!-\!m^2)(s_{134}\!-\!m^2)\dots(s_{13\dots(n-1)}\!-\!m^2)\,
             [3\;\!4]\braket{3|1|3\!+\!4|5}\braket{5\;\!6}\dots\braket{n\!-\!1|n} } . \nn
\end{align}
Here the $q$-dependent factors explicitly canceled after using \eqn{BCFWshift34kinem13},
and we also simplified
\be
   [1^a|\!-\!\hat{P}^c] \braket{\hat{P}_c\;\!2^b}
    =-[1^a|\hat{P}\:\!\ket{2^b}
    = \frac{m}{\bra{3}1|4]}
      \big( \braket{1^a 2^b} \bra{3}1|4]
          + \braket{1^a 3} \braket{2^b 3} [3\;\!4] \big) .
\ee
The contribution~\eqref{BCFWshift34z13} from $z_{13}$
in fact coincides with the $k=4$ term of the $\sum_{k=4}^{n-1}$ sum
in the full formula~\eqref{QQggnOM}. To see that, we only need to rewrite
\beal
   \frac{ \braket{3|\!\!\not{\!p}_{1}\!\not{\!p}_{34}|3} }
        { s_{34} \braket{34} \braket{3|\!\!\not{\!p}_{1}\!\not{\!p}_{34}|4} }
   \big( \braket{1^a 2^b} \braket{3|\!\!\not{\!p}_{1}\!\not{\!p}_{34}|3} &
       + \braket{1^a 3} \braket{2^b 3} s_{34} \big) \bra{3}\!\!\not{\!p}_{34} \\
 =-\frac{ \bra{3}1|4] }{ (s_{13}-m^2) [3\;\!4] }
   \big( \braket{1^a 2^b} \bra{3}1|4] &
       + \braket{1^a 3} \braket{2^b 3} [3\;\!4] \big) [4| .
\eeal

\begin{figure}[t]
\centering
\vspace{-7pt}
$ \scalegraph{1.0}{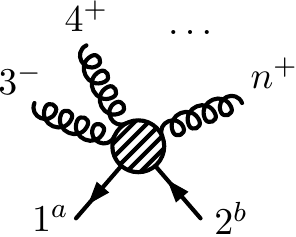}~=~\scalegraph{1.0}{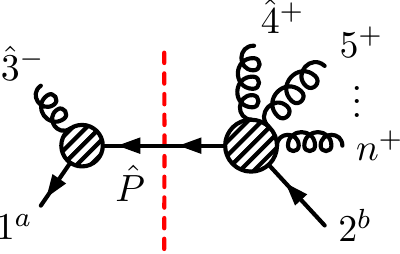}~+~~\scalegraph{1.0}{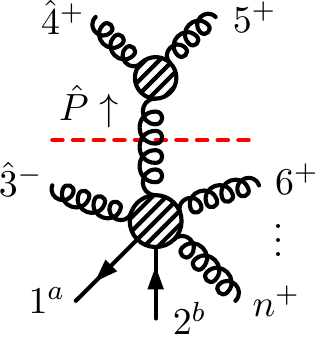} $
\vspace{-5pt}
\caption{\small Graphic representation of the BCFW recursion step~\eqref{bcfwOM}
for $A(\u{1}^a\:\!\!,3^-\!,4^+\!,\dots,n^+\!,\o{2}^b)$}
\label{fig:bcfwOM}
\end{figure}

Now we turn to the residue~\eqref{bcfwOM45} at $z_{45}$.
First of all, we observe that the right-hand three-gluon amplitude is invariably
\be
   \frac{-i}{s_{45}} A(-\hat{P}^-\!,\hat{4}^+\!,5^+)
    = \frac{-i}{s_{45}} \frac{-i[4\;\!5]^4}{[-\hat{P}|4][4\;\!5][5|\!-\!\hat{P}]}
    =-\frac{[4\;\!5]^2}{\braket{4\;\!5}[4\;\!\hat{P}][5\;\!\hat{P}]}
    = \frac{\braket{3\;\!5}}{\braket{3\;\!4}\braket{4\;\!5}} .
\ee
Furthermore, in the left-hand amplitude of \eqn{bcfwOM45},
$A(\u{1}^a\:\!\!,\hat{3}^-\!,\hat{P}^+\!,6^+\!,\dots,n^+\!,\o{2}^b)$,
most momentum sums (and their squares) are simply shifted:
\be
   p_{3\dots(k-1)} \to \hat{p}_{3P6 \dots k}
    = \hat{p}_3 + \hat{p}_4 + p_5 + p_6 + \ldots + p_k
    = p_{3 \dots k} , \qquad \quad
   s_{3\dots(k-1)} \to s_{3 \dots k} .
\ee
We can then compute the residue at $z_{45}$ as
\begin{subequations}\begin{align}
   A(&\u{1}^a\:\!\!,\hat{3}^-\!,\hat{P}^+\!,6^+\!,\dots,n^+\!,\o{2}^b)
      \frac{-i}{s_{45}} A(-\hat{P}^-\!,\hat{4}^+\!,5^+) \nn \\*
  = & \Bigg\{\!
    - \frac{ i \braket{3|1|2|3}\,
             \big( \braket{1^a 3} [2^b|1\!+\!2\ket{3}
                 + \braket{2^b 3} [1^a|1\!+\!2\ket{3} \big) }
           { s_{12} \braket{3\;\!\hat{P}} \braket{\hat{P}\;\!6} \braket{6\;\!7} \dots
             \braket{n\!-\!1|n} \braket{3|1|1\!+\!2|n} } \nn \\* &~\;
\label{BCFWshift34z45a}
    + \frac{ i\;\!m \braket{3|\!\!\not{\!p}_{1}\!\not{\!p}_{345}|3}\,
             \big( \braket{1^a 2^b}
                   \braket{3|\!\!\not{\!p}_{1}\!\not{\!p}_{345}|3}
                 + \braket{1^a 3} \braket{2^b 3} s_{345} \big) }
           { s_{345} (s_{1345}\!-\!m^2)\dots(s_{13\dots(n-1)}\!-\!m^2)\,
             \braket{3\;\!\hat{P}} \braket{3|\!\!\not{\!p}_{1}\!\not{\!p}_{345}|\hat{P}} } \\* &
      \qquad\:\times
      \frac{ \bra{3}\!\!\not{\!p}_{345}
             \big\{\!\!\not{\!p}_{1345}\!\not{\!p}_{6}+(s_{1345}-m^2) \big\}
             \prod_{j=6}^{n-2}\!\big\{\!\!\not{\!p}_{13 \dots j}\!\not{\!p}_{j+1}
                                                +(s_{13 \dots j}-m^2) \big\}|n] }
           { \braket{3|\!\!\not{\!p}_{1}\!\not{\!p}_{345}|6}
             \braket{6\;\!7}\dots\braket{n\!-\!1|n} } \nn \\ &~\;
    + \sum_{k=6}^{n-1}
      \frac{ i\;\!m \braket{3|\!\!\not{\!p}_{1}\!\not{\!p}_{3 \dots k}|3}\,
             \big( \braket{1^a 2^b}
                   \braket{3|\!\!\not{\!p}_{1}\!\not{\!p}_{3 \dots k}|3}
                 + \braket{1^a 3} \braket{2^b 3} s_{3 \dots k} \big) }
           { s_{3 \dots k} (s_{13\dots k}\!-\!m^2)\dots(s_{13\dots(n-1)}\!-\!m^2)\,
             \braket{3\;\!\hat{P}} \braket{\hat{P}\;\!6}
             \braket{6\;\!7} \dots \braket{k\!-\!1|k}
             \braket{3|\!\!\not{\!p}_{1}\!\not{\!p}_{3 \dots k}|k} } \nn \\* &
      \qquad\:\times
      \frac{ \bra{3}\!\!\not{\!p}_{3 \dots k}|
             \prod_{j=k}^{n-2}\!\big\{\!\!\not{\!p}_{13 \dots j}\!\not{\!p}_{j+1}
                                                +(s_{13 \dots j}-m^2) \big\}|n] }
           { \braket{3|\!\!\not{\!p}_{1}\!\not{\!p}_{3 \dots k}|k\!+\!1}
             \braket{k\!+\!1|k\!+\!2}\dots\braket{n\!-\!1|n} } \Bigg\}
      \frac{\braket{3\;\!5}}{\braket{3\;\!4}\braket{4\;\!5}} \nn \\
\label{BCFWshift34z45b}
   =&-\frac{ i \braket{3|1|2|3}\,
             \big( \braket{1^a 3} [2^b|1\!+\!2\ket{3}
                 + \braket{2^b 3} [1^a|1\!+\!2\ket{3} \big) }
           { s_{12} \braket{3\;\!4} \braket{4\;\!5} \braket{5\;\!6}
             \braket{6\;\!7} \dots \braket{n\!-\!1|n} \braket{3|1|1\!+\!2|n} } \\ &
    + \sum_{k=5}^{n-1}
      \frac{ i\;\!m \braket{3|\!\!\not{\!p}_{1}\!\not{\!p}_{3 \dots k}|3}\,
             \big( \braket{1^a 2^b}
                   \braket{3|\!\!\not{\!p}_{1}\!\not{\!p}_{3 \dots k}|3}
                 + \braket{1^a 3} \braket{2^b 3} s_{3 \dots k} \big) }
           { s_{3 \dots k} (s_{13\dots k}\!-\!m^2)\dots(s_{13\dots(n-1)}\!-\!m^2)\,
             \braket{3\;\!4}\braket{4\;\!5} \braket{5\;\!6}
             \braket{6\;\!7} \dots \braket{k\!-\!1|k}
             \braket{3|\!\!\not{\!p}_{1}\!\not{\!p}_{3 \dots k}|k} } \nn \\ &
      \qquad \qquad \qquad \qquad \qquad \qquad \qquad \quad~~~\;\times
      \frac{ \bra{3}\!\!\not{\!p}_{3 \dots k}
             \prod_{j=k}^{n-2}\!\big\{\!\!\not{\!p}_{13 \dots j}\!\not{\!p}_{j+1}
                                                +(s_{13 \dots j}-m^2) \big\}|n] }
           { \braket{3|\!\!\not{\!p}_{1}\!\not{\!p}_{3 \dots k}|k\!+\!1}
             \braket{k\!+\!1|k\!+\!2}\dots\braket{n\!-\!1|n} } .\! \nn
\end{align} \label{BCFWshift34z45}%
\end{subequations}
Here we were able to integrate the second term in the bracket~\eqref{BCFWshift34z45a},
which corresponds to $k=\hat{P}$,
into the $\sum_{k=6}^{n-1}$ sum as that for $k=5$.
Since the $k=4$ term, missing from \eqn{BCFWshift34z45},
is provided by the residue at $z_{13}$,
this concludes the proof of the formula~\eqref{QQggnOM}.

\section{Checks}
\label{sec:checks}

As the first simple check of our
all-multiplicity formulae~\eqref{QQggnAP} and \eqref{QQggnOM},
we evaluate their massless limits.
The former explicitly vanishes, as it should,
whereas the latter reduces to the massless MHV amplitudes with two quarks:
\be
   A(\u{1}^a\:\!\!,3^-\!,4^+\!,\dots,n^+\!,\o{2}^b) \xrightarrow[m\to0]{}
      \frac{i}{\braket{1\;\!3}\braket{3\;\!4}\braket{4\;\!5}\dots
               \braket{n\!-\!1|n}\braket{n|2}\braket{2\;\!1}}
      \begin{pmatrix}~\,0 \!\!&\!\! \braket{1\;\!3} \braket{2\;\!3}^3 \\
                     -\braket{1\;\!3}^3 \braket{2\;\!3} \!\!&\!\! 0 \end{pmatrix}\!.
\ee
This analytic check, however, is only sensitive to a single term in \eqn{QQggnOM}
that is not multiplied by the mass.
As another partial check,
we happened to have a six-point Feynman-diagrammatic calculation at easy access,
with which we found numerical agreement to ten significant digits
for both helicity configurations.
Needless to say, the Feynman diagrams were much lengthier before evaluation
than the three-term amplitude generated by the formula~\eqref{QQggnOM}.
The all-plus formula~\eqref{QQggnAP} can also be independently verified
via the fundamental BCJ relation~\cite{Bern:2008qj,Feng:2010my,Johansson:2015oia,delaCruz:2015dpa}
\be
   \sum_{i=2}^{n-1} (p_{13 \dots i}\cdot p_n)
      A(\u{1}^a\:\!\!,3^+\!,\dots,i^+\!,n^+\!,(i\!+\!1)^+\!,\dots,(n-1)^+\!,\o{2}^b) = 0 ,
\label{fBCJ}
\ee
which is non-trivially satisfied by a linear combination of its permutations.
We checked it numerically to ten significant digits for $n=4,5,\dots,12$.
Let us now turn to even more stringent checks.

In \rcite{Schwinn:2007ee} Schwinn and Weinzierl used
a massive extension~\cite{Kleiss:1986qc,Dittmaier:1998nn,Schwinn:2005pi}
of the massless spinor-helicity formalism
to compute QCD amplitudes with the same gluon polarizations that we compute here.
In that formalism, the massive spinors are introduced
by expanding the massive momentum~$p_\mu$ in terms of
its massless ``flat'' version~$p_\mu^\flat$ and another null vector~$q_\mu$:
\be
   p_{\mu} = p_{\mu}^{\flat} + \frac{m^{2}}{2(p\!\cdot\!q)} q_{\mu} , \qquad \quad
   \begin{aligned}
   u_{p}^+(q) = \frac{(\not{\!p}+m)\ket{q}}{\braket{p^\flat q}}, \qquad \quad
   u_{p}^-(q) = \frac{(\not{\!p}+m)|q]}{[p^\flat q]} , \\
   v_{p}^-(q) = \frac{(\not{\!p}-m)\ket{q}}{\braket{p^\flat q}}, \qquad \quad
   v_{p}^+(q) = \frac{(\not{\!p}-m)|q]}{[p^\flat q]} .
   \end{aligned}
\label{massivespinorsold}
\ee
The reference momentum $q_\mu$ determines the spin quantization axis, since
the resulting spin vector is
$\vec{\:\!s}^{\,\pm}=\pm\big\{\:\!^{\hat{p}}\!/\!_{2}\!
                     -\:\!^{m^2 \vec{\:\!q}}\!/\!_{4|\vec{\:\!p}|p \cdot q}\!\big\}$.
This shows that unless $m \neq 0$ the quark labels~``$\pm$''
in the spinors~\eqref{massivespinorsold}
are not helicities but rather general spin labels.

The versatility of the massive spinor-helicity formalism~\cite{Arkani-Hamed:2017jhn}
reviewed in \sec{sec:spinors} lets us effortlessly pick any quantization axis in the sense of \eqn{massivespinorsold}. We simply need to switch
from the helicity parametrization~\eqref{massivespinorsolutionangles} to
\beal
   \bar{u}_{p}^{a=1} &
    = \begin{pmatrix} -\bra{p^{1}} & \equiv & \frac{m\bra{q}}{\braket{q\;\!p^\flat}} \\
                         ~~[p^{1}| & \equiv & [p^{\flat}|
      \end{pmatrix} = \bar{u}_{p}^-(q) , \qquad \qquad\:\,
   v_p^{a=1}
    = \begin{pmatrix}-\ket{p^{1}} & \equiv &-\frac{m\ket{q}}{\braket{p^\flat q}} \\
                        ~~|p^{1}] & \equiv & |p^{\flat}]
      \end{pmatrix} = v_{p}^-(q) , \\
   \bar{u}_{p}^{a=2} &
    = \begin{pmatrix} -\bra{p^{2}} & \equiv &-\bra{p^{\flat}} \\
                         ~~[p^{2}| & \equiv &-\frac{m[q|}{[q\;\!p^\flat]}
      \end{pmatrix} =-\bar{u}_{p}^+(q) ,  \qquad \quad
   v_p^{a=2}
    = \begin{pmatrix}-\ket{p^{2}} & \equiv &-\ket{p^{\flat}} \\
                        ~~|p^{2}] & \equiv & \frac{m|q]}{[p^\flat q]}
      \end{pmatrix} =-v_{p}^+(q) .
\label{massivespinors2old}
\eeal
Now if we take the same reference vector $q^\mu$ for both quarks, this dictionary gives us
\be
   \braket{1^{a}\o{2}^{b}}
    = \begin{pmatrix} 0 & -m \frac{\braket{q\;\!2^\flat}}{\braket{q\;\!1^\flat}} \\
      m \frac{\braket{1^\flat q}}{\braket{2^\flat q}} & \braket{1^\flat 2^\flat}
      \end{pmatrix}\!, \qquad \quad
   \braket{1^a 3} \braket{2^b 3}
    = \begin{pmatrix}
      \frac{m^2 \braket{q\;\!3}^2}{\braket{q\;\!1^\flat} \braket{q\;\!2^\flat}} &
      -\frac{m\braket{q\;\!3}}{\braket{q\;\!1^\flat}} \braket{2^\flat 3} \\
      -\braket{1^\flat 3} \frac{m\braket{q\;\!3}}{\braket{q\;\!2^\flat}} &
      \braket{1^\flat 3} \braket{2^\flat 3}
      \end{pmatrix}\!.
\label{massivespinors2old2}
\ee

\Rcite{Schwinn:2007ee} actually sets $q^\mu$ to the momentum of the minus-helicity gluon~$3$.
This allowed for BCFW shifts
involving this pair of massive and massless momenta, and thus set up a recursion
to compute the amplitudes in question.
Let us translate the results of \rcite{Schwinn:2007ee}
to the conventions of the present paper:
\begin{subequations} \begin{align}
\label{QQggnOMpp}
 & A(\u{1}^{1}\:\!\!,3^-\!,4^+\!,\dots,n^+\!,\o{2}^{1}) = 0 , \\
\label{QQggnOMpm}
 & A(\u{1}^{1}\:\!\!,3^-\!,4^+\!,\dots,n^+\!,\o{2}^{2})
    = \frac{\:\!\!\!\!\!\!-i\braket{2^\flat\:\!3} }
           { \braket{1^\flat\:\!3} \braket{34} \dots \braket{n\!-\!1|n} }
      \sum_{k=4}^{n}
      \frac{ \braket{3|\!\!\not{\!p}_{1}\!\not{\!p}_{3 \dots k}|3}^2 }
           { s_{3 \dots k} \braket{3|\!\!\not{\!p}_{1}\!\not{\!p}_{3 \dots k}|k} } \nn \\*
    & \qquad\;\times
      \Bigg\{  \delta_{k=n} + \delta_{k\neq n}
      \frac{ m^2 \braket{k|k\!+\!1}
             \bra{3}\!\!\not{\!p}_{3 \dots k}
             \prod_{j=k+1}^{n-1}\!\big\{(s_{13 \dots j}-m^2)
              - \!\not{\!p}_{j}\!\not{\!p}_{13 \dots j} \big\}|n] }
           { (s_{13\dots k}\!-\!m^2)\dots(s_{13\dots(n-1)}\!-\!m^2)
             \braket{3|\!\!\not{\!p}_{1}\!\not{\!p}_{3 \dots k}|k\!+\!1} } \Bigg\} , \\
\label{QQggnOMmp}
 & A(\u{1}^{2}\:\!\!,3^-\!,4^+\!,\dots,n^+\!,\o{2}^{1})
    = \frac{i\braket{1^\flat\:\!3} }
           { \braket{2^\flat\:\!3} \braket{34} \dots \braket{n\!-\!1|n} }
      \sum_{k=4}^{n}
      \frac{ \braket{3|\!\!\not{\!p}_{1}\!\not{\!p}_{3 \dots k}|3}^2 }
           { s_{3 \dots k} \braket{3|\!\!\not{\!p}_{1}\!\not{\!p}_{3 \dots k}|k} } \nn \\*
    & \qquad\;\times
      \Bigg\{  \delta_{k=n} + \delta_{k\neq n}
      \frac{ m^2 \braket{k|k\!+\!1}
             \bra{3}\!\!\not{\!p}_{3 \dots k}
             \prod_{j=k+1}^{n-1}\!\big\{(s_{13 \dots j}-m^2)
              - \!\not{\!p}_{j}\!\not{\!p}_{13 \dots j} \big\}|n] }
           { (s_{13\dots k}\!-\!m^2)\dots(s_{13\dots(n-1)}\!-\!m^2)
             \braket{3|\!\!\not{\!p}_{1}\!\not{\!p}_{3 \dots k}|k\!+\!1} } \Bigg\} , \\
\label{QQggnOMmm}
 & A(\u{1}^{2}\:\!\!,3^-\!,4^+\!,\dots,n^+\!,\o{2}^{2})
    = \frac{i \braket{1^\flat\:\!2^\flat}}{m \braket{34} \dots \braket{n\!-\!1|n}}
      \sum_{k=4}^{n}
      \frac{ \braket{3|\!\!\not{\!p}_{1}\!\not{\!p}_{3 \dots k}|3}^2 }
           { s_{3 \dots k} \braket{3|\!\!\not{\!p}_{1}\!\not{\!p}_{3 \dots k}|k} }
      \bigg[ 1 + \frac{ s_{3 \dots k}\braket{3\;\!2^\flat} }
                      { \braket{3|\!\!\not{\!p}_{3 \dots k}\!\not{\!p}_{1}^\flat|2^\flat} }
      \bigg] \nn \\*
    & \qquad\;\times
      \Bigg\{  \delta_{k=n} + \delta_{k\neq n}
      \frac{ m^2 \braket{k|k\!+\!1}
             \bra{3}\!\!\not{\!p}_{3 \dots k}
             \prod_{j=k+1}^{n-1}\!\big\{(s_{13 \dots j}-m^2)
              - \!\not{\!p}_{j}\!\not{\!p}_{13 \dots j} \big\}|n] }
           { (s_{13\dots k}\!-\!m^2)\dots(s_{13\dots(n-1)}\!-\!m^2)
             \braket{3|\!\!\not{\!p}_{1}\!\not{\!p}_{3 \dots k}|k\!+\!1} } \Bigg\} .
\end{align} \label{QQggnOMsw}%
\end{subequations}
To make a direct comparison easier, here we rewrite our result~\eqref{QQggnOM} as
\beal\!\!
   A(\u{1}^a\:\!\!,3^-\!,4^+\!,\dots,n^+\!,\o{2}^b) \qquad \qquad \qquad
      \qquad \qquad \qquad \qquad \qquad \qquad \qquad \qquad \qquad \quad~\;& \\
    = \frac{i}{\braket{34} \dots \braket{n\!-\!1|n}}
      \Bigg\{
      \frac{ \braket{3|\!\!\not{\!p}_{1}\!\not{\!p}_{3 \dots n}|3} }
           { m\:\!s_{3 \dots n} \braket{3|\!\!\not{\!p}_{1}\!\not{\!p}_{3 \dots n}|n} }
      \big( \braket{1^a 2^b}
            \braket{3|\!\!\not{\!p}_{1}\!\not{\!p}_{3 \dots n}|3}
          + \braket{1^a 3} \braket{2^b 3} s_{3 \dots n} \big)~\;\,& \\
    +\,m \sum_{k=4}^{n-1}
      \frac{ \braket{k|k\!+\!1} \braket{3|\!\!\not{\!p}_{1}\!\not{\!p}_{3 \dots k}|3}
             \bra{3}\!\!\not{\!p}_{3 \dots k}
             \prod_{j=k}^{n-2}\!\big\{ (s_{13\dots(j+1)}\!-m^2)\,
              -\!\not{\!p}_{j+1}\!\not{\!p}_{13\dots(j+1)} \big\}|n] }
           { s_{3 \dots k}\,(s_{13\dots k}\!-\!m^2)\dots(s_{13\dots(n-1)}\!-\!m^2)\,
             \braket{3|\!\!\not{\!p}_{1}\!\not{\!p}_{3 \dots k}|k}
             \braket{3|\!\!\not{\!p}_{1}\!\not{\!p}_{3 \dots k}|k\!+\!1} } & \\ \times
             \big( \braket{1^a 2^b}
                   \braket{3|\!\!\not{\!p}_{1}\!\not{\!p}_{3 \dots k}|3}
                 + \braket{1^a 3} \braket{2^b 3} s_{3 \dots k} \big) & \Bigg\} ,\!\!\!
\label{QQggnOMalt}
\eeal
where we massaged the product $\prod_{j=k}^{n-2}$ into the form of \eqn{QQggnOMsw}
using the anticommutator identity
\be
   \not{\!p}_{13 \dots j}\!\not{\!p}_{j+1}+(s_{13 \dots j}-m^2)
    = (s_{13\dots(j+1)}-m^2)-\!\not{\!p}_{j+1}\!\not{\!p}_{13\dots(j+1)} .
\label{anticommutator13j}
\ee
Now we can see with a naked eye that our formula~\eqref{QQggnOMalt}
exactly reproduces \eqnss{QQggnOMpp}{QQggnOMpm}{QQggnOMmp},
where all the terms containing $\braket{1^a 3} \braket{2^b 3}$ vanish
due to the choice $q=p_3$ for both quarks.
To match the last amplitude~\eqref{QQggnOMmm},
for which $\ket{1^{a=2}}=\ket{1^\flat}$, $\ket{2^{b=2}}=\ket{2^\flat}$,
we observe that
\be
   \braket{1^\flat\:\!2^\flat} \braket{3|\!\!\not{\!p}_{1}\!\not{\!p}_{3 \dots k}|3}
    + \braket{1^\flat\:\!3} \braket{2^\flat\:\!3} s_{3 \dots k}
    = \braket{1^\flat\:\!2^\flat} \braket{3|\!\!\not{\!p}_{1}\!\not{\!p}_{3 \dots k}|3}
      \bigg[ 1 + \frac{ s_{3 \dots k}\braket{3\;\!2^\flat} }
                      { \braket{3|\!\!\not{\!p}_{3 \dots k}\!\not{\!p}_{1}^\flat|2^\flat} }
      \bigg] .
\ee

To conclude, we note that \rcite{Schwinn:2007ee} also computed
the analogue of the all-plus amplitude
in the massless spinor-helicity formalism~\eqref{massivespinorsold}
through its relation to the massive scalar amplitude of \rcites{Forde:2005ue,Ferrario:2006np}
via a supersymmetric Ward identity~\cite{Schwinn:2006ca,Craig:2011ws,Boels:2011zz}
with an unfixed reference vector $q^\mu$.
The same \eqns{massivespinors2old2}{anticommutator13j} allow to easily verify
that these results are incorporated in our formula~\eqref{QQggnAP}.

\section{Summary and discussion}
\label{sec:outro}

In this note we have computed two infinite families of tree-level amplitudes with two quarks of arbitrary spin and any number of gluons with specified helicities.
For that we have used
the new massive spinor-helicity formalism of \rcite{Arkani-Hamed:2017jhn}.
In order to check the consistency of our results with the literature,
we have also established straightforward transition rules
between our approach and the more traditional ones.

We hope to have demonstrated that
the new massive formalism is a analytic tool well-suited for QCD computations.
It is a logical extension of the massless spinor-helicity formalism~\cite{Berends:1981rb,DeCausmaecker:1981bg,Gunion:1985vca,Kleiss:1985yh,Xu:1986xb,Gastmans:1990xh,Arkani-Hamed:2017jhn},
which in the last decades has become indispensable for scattering-amplitude calculations.
Of course, the scope of the formalism is much more general than QCD,
as shown by the recent applications to gravitational scattering~\cite{Guevara:2017csg,Moynihan:2017tva}
and the Standard Model as a whole~\cite{Arkani-Hamed:2017jhn,Christensen:2018zcq}.
It can be used streamline the consideration of
all unitarity-compliant three-point~\cite{Conde:2016vxs,Conde:2016izb,Christensen:2018zcq}
and four-point~\cite{Arkani-Hamed:2017jhn} interactions.
It can also be related to much earlier
off-shell reformulations of QED~\cite{Brown:1958zz,Tonin:1959}
and other theories~\cite{Chalmers:1997ui,Chalmers:1998jb,Chalmers:2001cy}
using two-component spinor fields.

The presented formalism has potential to facilitate many QCD calculations,
both analytically and numerically.
Through its analytic simplicity, it may provide a way to explicit expressions
for tree amplitudes with more general gluon helicity configurations~\cite{Badger:2005zh,Badger:2005jv,Ozeren:2006ft,Huang:2012gs} and more quark-antiquark lines~\cite{Johansson:2014zca,Johansson:2015oia},
as already achieved~\cite{Drummond:2008cr,Dixon:2010ik}
for the massless QCD amplitudes with up to three quark-antiquark pairs. For example,
it would be interesting to find an analytic expression even for an amplitude
with two quarks and one negative-helicity gluon in an arbitrary position,
provided that it is more compact than its BCJ relation~\cite{Bern:2008qj,Johansson:2015oia}
that involves various permutations of the formula computed in this note.

New loop amplitudes could also be calculated using the presented formalism.
Indeed, loops can be obtained from generalized unitarity cuts~\cite{Bern:1994zx,Bern:1994cg,%
Britto:2004nc,Forde:2007mi,Britto:2005ha,Anastasiou:2006jv,Giele:2008ve}
that are constructed from tree amplitudes.\footnote{A relevant example
is the recent calculation~\cite{Badger:2017gta}
of one-loop amplitudes with external massive fermions
via tree amplitudes in six dimensions.
One extra dimension was needed
to parametrize the loop dependence in dimensional regularization,
and another one to account for the quark mass.
Since the massive four-dimensional spinor helicity
can be viewed as massless but five-dimensional,
the amplitudes considered in this note could be embedded
into a higher-dimensional formalism,
such as the six-dimensional one of \rcite{Cheung:2009dc}.}
It would also be interesting to investigate,
in the spirit of \rcites{Dinsdale:2006sq,Badger:2012uz},
if the massive on-shell formalism
could speed up numerical evaluation of tree-level amplitudes.
This would be beneficial for computing real-emission radiative QCD corrections
to a vast array of elementary-particle scattering processes.

\begin{acknowledgments}

The author would like to thank Nikolay Gulitskiy and Achilleas Lazopoulos
for their valuable comments on the manuscript,
as well as Canxin Shi for helping with the switch to the present conventions.
This project has received funding from the European Union's Horizon 2020 research and innovation programme under the Marie Sk{\l}odowska-Curie grant agreement 746138.

\end{acknowledgments}

\appendix

\section{Massless spinor parametrizations}
\label{sec:masslessspinors}

For this note to be more self-contained, we give a realization
of the massless spinors~\cite{Berends:1981rb,DeCausmaecker:1981bg,
Gunion:1985vca,Kleiss:1985yh,Xu:1986xb,Gastmans:1990xh}.
In terms of the light-cone momentum components
\be
   p_{+} = p^0 + p^3 , \qquad \quad
   p_{-} = p^0 - p^3 , \qquad \quad
   p_{\perp} = p^1 + ip^2 , \qquad \quad
   \o{p}_{\perp} = p^1 - ip^2 ,
\label{momentumlightcone}
\ee
satisfying $ p_{+} p_{-} = p_{\perp} \o{p}_{\perp} $,
an explicit solution for the Weyl spinors is
(adapted from \rcite{Maitre:2007jq})
\be
   \la_{p\:\!\alpha} = i^{\theta(-p^0)}
      \begin{pmatrix}
         ^{-\o{p}_{\perp}}\!/\!_{\sqrt{p_{+}}} \\
         \sqrt{p_{+}}
      \end{pmatrix}\!,
   \qquad \quad
   \lb_{p\:\!\dot{\alpha}} = (-i)^{\theta(-p^0)}
      \begin{pmatrix}
         ^{-p_{\perp}}\!/\!_{\sqrt{p_{+}}} \\
         \sqrt{p_{+}}
      \end{pmatrix}\!.
\label{spinorsolution1}
\ee
In the case of a real-valued momentum with positive energy,
it is just a rewrite of \eqn{spinorsolutionangles}.
For negative energies, the step functions $\theta(-p^0)$
introduce the $\pm i$ prefactors to ensure
the momentum inversion rule
\be
   \la_{-p} = -\la_{p} ,
   \qquad \quad
   \lb_{-p} = \lb_{p} ,
\label{momentuminversion}
\ee
assuming the principal square roots.
Moreover, for a real-valued momentum $p^\mu$
the spinor conjugation property is
$(\la_{p\:\!\alpha})^*=\sgn(p^0)\lb_{p\:\!\dot{\alpha}}$.
If $p_{+}$ happens to vanish, an equivalent solution may be used:
\be
   \la_{p\:\!\alpha} = i^{\theta(-p^0)}
      \begin{pmatrix}
         -\sqrt{p_{-}} \\
         ^{p_{\perp}}\!/\!_{\sqrt{p_{-}}}
      \end{pmatrix}\!,
   \qquad \quad
   \lb_{p\:\!\dot{\alpha}} = (-i)^{\theta(-p^0)}
      \begin{pmatrix}
         -\sqrt{p_{-}} \\
         ^{\o{p}_{\perp}}\!/\!_{\sqrt{p_{-}}}
      \end{pmatrix}\!.
\label{spinorsolution2}
\ee
In the complex-valued case where $p_{\pm}=0$
and the momentum equals $(0,p^1,\pm ip^1,0)$,
a valid choice is
\be
   \la_{p\:\!\alpha} = \frac{1}{\sqrt{2p^1}}
      \begin{pmatrix}
         -\o{p}_{\perp} \\
         p_{\perp}
      \end{pmatrix}\!,
   \qquad \quad
   \lb_{p\:\!\dot{\alpha}} = \frac{1}{\sqrt{2p^1}}
      \begin{pmatrix}
         -p_{\perp} \\
         \o{p}_{\perp}
      \end{pmatrix}\!.
\label{spinorsolution3}
\ee

\section{Massive spinor parametrizations}
\label{sec:massivespinors}

Here we give the massive spinor-helicity variables that are consistent
with the parametrizations~\eqref{spinorsolution1} through \eqref{spinorsolution3}
in the massless limit.
The spinors~\eqref{massivespinorsolutiondecomposition} can be rewritten as
\begin{subequations}
\begin{align}
   \la_{p\:\!\alpha}^{~\;a} =
      i^{\theta(-E)} & \Bigg\{
      \sqrt{\frac{E\!+\!P}{2P}}
      \begin{pmatrix} ^{-\o{p}_{\perp}}\!/\!_{\sqrt{p_{+}}} \\ \sqrt{p_{+}}
      \end{pmatrix}_{\!\alpha}\!\!\!\otimes\!
      \begin{pmatrix} 0 \\ 1
      \end{pmatrix}^{\!a}\!
    + \sqrt{\frac{E\!-\!P}{2P}}
      \begin{pmatrix} \sqrt{p_{+}} \\ ^{p_{\perp}}\!/\!_{\sqrt{p_{+}}}
      \end{pmatrix}_{\!\alpha}\!\!\!\otimes\!
      \begin{pmatrix} 1 \\ 0
      \end{pmatrix}^{\!a}\!\Bigg\} , \\
   \lb_{p\:\!\dot{\alpha}}^{~\;a} =
      (-i)^{\theta(-E)} & \Bigg\{
      \sqrt{\frac{E\!+\!P}{2P}}
      \begin{pmatrix} ^{-p_{\perp}}\!/\!_{\sqrt{p_{+}}} \\ \sqrt{p_{+}}
      \end{pmatrix}_{\!\dot{\alpha}}\!\!\!\otimes\!
      \begin{pmatrix} 1 \\ 0
      \end{pmatrix}^{\!a}\!
    - \sqrt{\frac{E\!-\!P}{2P}}
      \begin{pmatrix} \sqrt{p_{+}} \\ ^{\o{p}_{\perp}}\!/\!_{\sqrt{p_{+}}}
      \end{pmatrix}_{\!\dot{\alpha}}\!\!\!\otimes\!
      \begin{pmatrix} 0 \\ 1
      \end{pmatrix}^{\!a}\! \Bigg\} .
\end{align}
\label{massivespinorsolution1}%
\end{subequations}
where now we take $p_{\pm} = P \pm p^3$, $p_{\perp} = p^1 + ip^2$.
Moreover, we introduce a sign function in the definition
\be
   P = \sgn(E)\sqrt{\vec{\:\!p}^2} .
\label{pmsigns}
\ee
This allows the massless limit
to keep $(E+P)$ finite and send $(E-P)\to m^2/(2P)$
for all real-valued energies $E=p^0$,
as well as preserve the sign of
$\det\{\la_{p\:\!\alpha}^{~\;a}\}=\det\{\lb_{p\:\!\dot{\alpha}}^{~\;a}\}=m>0$
at the same time.
If $p_{+}$ happens to vanish, an equivalent solution may be used:\footnote{Using that
$\sqrt{p_+}=\sqrt{2P}\cos({\theta}/{2})$,
$\sqrt{p_-}=\sqrt{2P}\sin({\theta}/{2})$ and
$p_\perp=2Pe^{i\varphi}\sin({\theta}/{2})\cos({\theta}/{2})$,
one can relate the parametrizations~\eqref{massivespinorsolution1}
and~\eqref{massivespinorsolution2} by the spin-preserving little-group rotation
$ \omega^{a}_{~b}
= \Big(\begin{smallmatrix}\!e^{-i\varphi}\!& 0 \\ 0 &\!e^{i\varphi}\!\end{smallmatrix}\Big) $.
}
\begin{subequations}
\begin{align}
   \la_{p\:\!\alpha}^{~\;a} =
      i^{\theta(-E)} & \Bigg\{
      \sqrt{\frac{E\!+\!P}{2P}}
      \begin{pmatrix} -\sqrt{p_{-}} \\ ^{p_{\perp}}\!/\!_{\sqrt{p_{-}}}
      \end{pmatrix}_{\!\alpha}\!\!\!\otimes\!
      \begin{pmatrix} 0 \\ 1
      \end{pmatrix}^{\!a}\!
    + \sqrt{\frac{E\!-\!P}{2P}}
      \begin{pmatrix} ^{\o{p}_{\perp}}\!/\!_{\sqrt{p_{-}}} \\ \sqrt{p_{-}} 
      \end{pmatrix}_{\!\alpha}\!\!\!\otimes\!
      \begin{pmatrix} 1 \\ 0
      \end{pmatrix}^{\!a}\!\Bigg\} , \\*
   \lb_{p\:\!\dot{\alpha}}^{~\;a} =
      (-i)^{\theta(-E)} & \Bigg\{
      \sqrt{\frac{E\!+\!P}{2P}}
      \begin{pmatrix} -\sqrt{p_{-}} \\ ^{\o{p}_{\perp}}\!/\!_{\sqrt{p_{-}}}
      \end{pmatrix}_{\!\dot{\alpha}}\!\!\!\otimes\!
      \begin{pmatrix} 1 \\ 0
      \end{pmatrix}^{\!a}\!
    - \sqrt{\frac{E\!-\!P}{2P}}
      \begin{pmatrix} ^{p_{\perp}}\!/\!_{\sqrt{p_{-}}} \\ \sqrt{p_{-}} 
      \end{pmatrix}_{\!\dot{\alpha}}\!\!\!\otimes\!
      \begin{pmatrix} 0 \\ 1
      \end{pmatrix}^{\!a}\!\Bigg\}  .
\end{align}
\label{massivespinorsolution2}%
\end{subequations}
Another singular region is where $P=0$, and thus $p^0=\pm m$ and $p_{\pm}=\pm p^3$.
Then we choose\!\!\!
\begin{subequations}
\begin{align} \!\!\!\!
   \la_{p\:\!\alpha}^{~\;a} =
      i^{\theta(-p^0)} \Bigg\{\!
      \sqrt{\o{p}_\perp\!}
      \begin{pmatrix} \sqrt{^{(p^0-p_+)}\!/\!_{p_\perp}\!}\, \\
                     -\sqrt{\;\!^{p_\perp}\!/\!_{(p^0-p_+)}\!}\,
      \end{pmatrix}_{\!\alpha}\!\!\!\!\otimes\!
      \begin{pmatrix} 0 \\ 1
      \end{pmatrix}^{\!a}\!\!
    + \frac{p^0}{\sqrt{p^0\!-p_+}}
      \begin{pmatrix} 0 \\ -\sqrt{\,^{p_{\perp}}\!/_{\o{p}_\perp}\!}\,
      \end{pmatrix}_{\!\alpha}\!\!\!\!\otimes\!
      \begin{pmatrix} 1 \\ 0
      \end{pmatrix}^{\!a}\!\Bigg\} , \!\! & \\ \!\!\!\!
   \lb_{p\:\!\dot{\alpha}}^{~\;a} =
      (-i)^{\theta(-p^0)} \Bigg\{\!
      \sqrt{p_\perp\!}
      \begin{pmatrix} \sqrt{^{(p^0-p_+)}\!/_{\o{p}_\perp}\!}\, \\
                     -\sqrt{\,^{\o{p}_\perp}\!/\!_{(p^0-p_+)}\!}\,
      \end{pmatrix}_{\!\dot{\alpha}}\!\!\!\!\otimes\!
      \begin{pmatrix} 1 \\ 0
      \end{pmatrix}^{\!a}\!\!
    - \frac{p^0}{\sqrt{p^0\!-p_+}}
      \begin{pmatrix} 0 \\ -\sqrt{\;\!^{\o{p}_{\perp}}\!/\!_{p_\perp}\!}\,
      \end{pmatrix}_{\!\dot{\alpha}}\!\!\!\!\otimes\!
      \begin{pmatrix} 0 \\ 1
      \end{pmatrix}^{\!a}\!\Bigg\} , \!\! &
\end{align}
\label{massivespinorsolution11}%
\end{subequations}
the massless limit of which is consistent with that of \eqn{massivespinorsolution1}
(evaluated at $P=0$).
These spinors do not, however, correspond to definite helicities,
as the spin vector defined by \eqn{spin3vector} turns out to be
$\vec{\:\!s}_{\:\!a}=(-1)^{a-1}\big\{\vec{\:\!p}
                                    -(0,0,\:\!^{m^2}\!/\!_{(p^0-p_+)})\big\}/(2p^0)$
for a complex $p^\mu$.
Similarly, in the case where $p_{\pm}=0$
and the momentum equals $(p^0,p^1,\pm ip^1,0)$, a valid choice is
\begin{subequations}
\begin{align}
   \la_{p\:\!\alpha}^{~\;a} =
      i^{\theta(-p^0)} & \Bigg\{
      \frac{1}{\sqrt{2p^1}}
      \begin{pmatrix} -\o{p}_{\perp}\!+p^0 \\ p_{\perp}\!-p^0\!\!
      \end{pmatrix}_{\!\alpha}\!\!\!\otimes\!
      \begin{pmatrix} 0 \\ 1
      \end{pmatrix}^{\!a}\!
    + \frac{1}{\sqrt{2p^1}}
      \begin{pmatrix} \mp p^0 \\ \pm p^0
      \end{pmatrix}_{\!\alpha}\!\!\!\otimes\!
      \begin{pmatrix} 1 \\ 0
      \end{pmatrix}^{\!a}\!\Bigg\} , \\
   \lb_{p\:\!\dot{\alpha}}^{~\;a} =
      (-i)^{\theta(-p^0)} & \Bigg\{
      \frac{1}{\sqrt{2p^1}}
      \begin{pmatrix} -p_{\perp}\!\mp p^0 \\ \o{p}_{\perp}\!\mp p^0\!\!
      \end{pmatrix}_{\!\dot{\alpha}}\!\!\!\otimes\!
      \begin{pmatrix} 1 \\ 0
      \end{pmatrix}^{\!a}\!
    - \frac{1}{\sqrt{2p^1}}
      \begin{pmatrix} 2p_{\perp}\!-p^0 \\ 2\o{p}_{\perp}\!-p^0
      \end{pmatrix}_{\!\dot{\alpha}}\!\!\!\otimes\!
      \begin{pmatrix} 0 \\ 1
      \end{pmatrix}^{\!a}\!\Bigg\} ,
\end{align}
\label{massivespinorsolution3}%
\end{subequations}
yielding the spin vectors
$\vec{\:\!s}_{\:\!a}=(-1)^{a-1}\big\{\:\!^{\vec{\:\!p}}\!/\!_{2p^0}\!
-(1,\pm i(1-\:\!^{p^0}\!/\!_{2p^1}\!),\pm (1-\:\!^{p^0}\!/\!_{2p^1}\!)\big\}$.
For a real massive momentum at rest $p^\mu=(p^0,0,0,0)$ we choose
\begin{subequations}
\begin{align}
   \la_{p\:\!\alpha}^{~\;a} =
      i^{\theta(-p^0)} & \Bigg\{
      \begin{pmatrix} 0 \\ \sqrt{p^0}
      \end{pmatrix}_{\!\alpha}\!\!\!\otimes\!
      \begin{pmatrix} 0 \\ 1
      \end{pmatrix}^{\!a}\!
    + \begin{pmatrix} \sqrt{p^0} \\ 0
      \end{pmatrix}_{\!\alpha}\!\!\!\otimes\!
      \begin{pmatrix} 1 \\ 0
      \end{pmatrix}^{\!a}\!\Bigg\} , \\
   \lb_{p\:\!\dot{\alpha}}^{~\;a} =
      (-i)^{\theta(-p^0)} & \Bigg\{
      \begin{pmatrix} 0 \\ \sqrt{p^0}
      \end{pmatrix}_{\!\dot{\alpha}}\!\!\!\otimes\!
      \begin{pmatrix} 1 \\ 0
      \end{pmatrix}^{\!a}\!
    - \begin{pmatrix} \sqrt{p^0} \\ 0
      \end{pmatrix}_{\!\dot{\alpha}}\!\!\!\otimes\!
      \begin{pmatrix} 0 \\ 1
      \end{pmatrix}^{\!a}\!\Bigg\} ,
\end{align}
\label{massivespinorsolution4}%
\end{subequations}
This choice naturally aligns the spin vector $\vec{\:\!s}^{\,a}=(-1)^{a-1}(0,0,1/2)$\
with the $z$-axis.
Finally, we note that for real-valued momenta $p^\mu$ the above definitions satisfy
$(\la_{p\:\!\alpha})^*=\sgn(p^0)\lb_{p\:\!\dot{\alpha}}$ and
$(\eta_{p\:\!\alpha})^*=\pm\tilde{\eta}_{p\:\!\dot{\alpha}}$
and are consistent with the momentum inversion rule~\eqref{momentuminversion}.

\bibliographystyle{JHEP}
\bibliography{references}

\end{document}